\documentclass[conference]{IEEEtran}
\usepackage{cite}
\usepackage{amsmath,amssymb,amsfonts}
\usepackage{algorithmic}
\usepackage{graphicx}
\usepackage{textcomp}
\usepackage{booktabs}
\usepackage{caption}
\usepackage{subcaption}
\usepackage{multirow}
\usepackage{comment}
\usepackage{url}
\usepackage{xcolor}

\IEEEoverridecommandlockouts

\definecolor{YellowOrange}{RGB}{255,127,0}
\definecolor{GreenYellow}{RGB}{255,127,0}
\definecolor{Cyan}{RGB}{255,127,0}
\definecolor{Mulberry}{RGB}{255,127,0}

\definecolor{stromaword}{RGB}{255, 150, 255}
\definecolor{otherword}{RGB}{0, 0, 255}
\definecolor{applegreen}{RGB}{0, 180, 30}

\newcommand{\arch}{HATNet}

\usepackage{tikz}
\usetikzlibrary{decorations.pathreplacing}
\usetikzlibrary{shapes.callouts}
\usetikzlibrary{chains, fit, quotes, automata}
\usetikzlibrary{arrows}
\usetikzlibrary{matrix,positioning}
\usetikzlibrary{shapes.geometric}
\usetikzlibrary{calc}

\def\BibTeX{{\rm B\kern-.05em{\sc i\kern-.025em b}\kern-.08em
    T\kern-.1667em\lower.7ex\hbox{E}\kern-.125emX}}
\begin{document}

\title{\arch: An End-to-End Holistic Attention Network for Diagnosis of Breast Biopsy Images
\thanks{This work was supported in part by the National Cancer Institute awards (R01 CA172343, R01 CA140560, and RO1 CA200690), NSF (IIS-1616112, IIS1252835), an Allen Distinguished Investigator award, and gifts from the Allen Institute for AI, Google, and Amazon.}
}

\author{\IEEEauthorblockN{
Sachin Mehta\IEEEauthorrefmark{1}\thanks{S. Mehta and X. Lu contributed equally.},
Ximing Lu\IEEEauthorrefmark{1},
Donald Weaver\IEEEauthorrefmark{2}, 
Joann G. Elmore\IEEEauthorrefmark{3},
Hannaneh Hajishirzi\IEEEauthorrefmark{1},
Linda Shapiro\IEEEauthorrefmark{1}
\\ \\
\IEEEauthorrefmark{1}University of Washington, Seattle, WA \\
\IEEEauthorrefmark{2}University of Vermont College of Medicine, Burlington, VT\\
\IEEEauthorrefmark{3}University of California, Los Angeles, CA
}}

\maketitle

\begin{abstract}
Training end-to-end networks for classifying gigapixel size histopathological images is computationally intractable. Most approaches are patch-based and first learn local representations (patch-wise) before combining these local representations to produce image-level decisions. However, dividing large tissue structures into patches limits the context available to these networks, which may reduce their ability to learn representations from clinically relevant structures. In this paper, we introduce a novel attention-based network, the Holistic ATtention Network (\arch) to classify breast biopsy images. We streamline the histopathological image classification pipeline and show how to learn representations from gigapixel size images end-to-end. \arch~extends the bag-of-words approach and uses self-attention to encode global information, allowing it to learn representations from clinically relevant tissue structures without any explicit supervision. It outperforms the previous best network Y-Net,  which uses supervision in the form of tissue-level segmentation masks, by 8\%.  Importantly, our analysis reveals that \arch~learns representations from clinically relevant structures, and it matches the classification accuracy of human pathologists for this challenging test set.
\end{abstract}

\begin{IEEEkeywords}
Self-attention, Transformers, Histopathological images, Breast cancer, Image classification
\end{IEEEkeywords}

\section{Introduction}
\label{sec:introduction}
\IEEEPARstart{B}{reast} cancer is  the most common cancer in women accounting for approximately 25\% of all cancer instances worldwide \cite{makki2015diversity,desantis2019breast}. Diagnostic classification errors among pathologists can have significant adverse consequences for patients. The ``gold standard" for diagnosis of breast biopsy specimens relies on a pathologist’s visual assessment of tissue sections and cognitive processing of learned cytologic and morphological criteria, including architectural and cellular changes in the tissue, alterations of the tumor micro-environment, and immune-mediated host response. Assessment of these morphological criteria is subjective and pathologists, even expert pathologists, cannot always reach consensus on diagnostically challenging cases. Diagnostic disagreement occurs throughout the spectrum of benign to malignant lesions \cite{elmore2015diagnostic}. Diagnostic variability is a serious problem as misclassifying breast cancer as benign may lead to fatal progression, and diagnosing a benign lesion as malignant may result in significant morbidity including overtreatment, unnecessary emotional strain, anxiety, and cost.  Additionally, misdiagnosis of breast cancer has been a leading cause for malpractice claims for decades, and misclassification undoubtedly undermines research quality and hinders medical progress \cite{kern2001delayed,reisch2015medical}. A computer-aided diagnostic system that could reduce classification uncertainties would have immediate positive clinical impact.

This paper introduces a self-attention-based network called {\bf H}olistic {\bf AT}tention {\bf Net}work (\arch) for classifying breast biopsy images in an end-to-end manner. \arch~extends the self-attention network of Vaswani et al. \cite{vaswani2017attention}. The core principle is to factorize the input biopsy image into words (or patches) using a bag-of-words approach and then encode inter-word and inter-bag relationships in a hierarchical manner using self-attention. Self-attention enables interaction between inputs (bags or words), allowing the encoding of global information in an end-to-end fashion. This helps the network learn clinically relevant tissue structures without any supervision.

\arch~significantly outperforms state-of-the-art methods; it is 8\% more accurate and about $2\times$ faster than the previous best network, Y-Net \cite{mehta2018ynet}, and also matches the classification performance of participant pathologists on the test set. Importantly, this network pays attention to ductal regions and stromal tissues, important bio-markers in breast cancer diagnosis, suggesting that there is clinical relevance in our method. To the best of our knowledge, this is the first work that (1) uses transformers to classify histopathological images in an end-to-end fashion and (2) correlates model decisions with clinically relevant structures. Our source code is available at \textcolor{blue}{\url{https://github.com/sacmehta/HATNet}}.

\section{Related Work}
\label{sec:related_work}

\noindent \textbf{Histopathological image classification:} Convolutional neural networks (CNNs) are state-of-the-art networks for image classification (e.g., ResNet \cite{he2016deep}), including histopathological image analysis \cite{cirecsan2013mitosis,xu2015deep,cruz2014automatic,hou2016patch,gecer2018detection,mehta2018ynet}. These methods follow a bag-of-words model for learning representations, wherein an image is a treated as a bag while image patches are treated as words (or instances). The first line of research focuses on extracting word-level representations using CNNs, which are then aggregated to produce image-level decisions. Feature selection-based aggregation methods (e.g., \cite{xu2015deep,cruz2014automatic,sun2019deep}) allows identification of relevant features in these word representations. However, such methods fail to capture the heterogeneity of diagnosis categories \cite{hou2016patch}. To address the limitations of these methods, multi-instance learning based methods have been proposed \cite{mercan2017multi,gecer2018detection,hou2016patch,wang2019weakly}. These methods first identify salient instances (or words), which are then combined using different methods (e.g., thresholding, majority voting, and learned fusion) to produce image-level decisions.

The second line of research considers tissue type, size, and distribution to produce image-level decisions \cite{mehta2018learning,mercan2019assessment,lu2015automated}. Instead of identifying salient words, these approaches produce word-level segmentation masks, which are then combined to produce image-level segmentation masks. Tissue-level structural information (e.g., size and distribution) extracted from these masks is then used to produce diagnosis categories.

The third line of research integrates both saliency-based and segmentation-based approaches \cite{mehta2018ynet,thome2019multitask,heker2020joint,hou2020multi}. These approaches simultaneously produce saliency maps and segmentation masks, which are then combined to extract structural information about tissues and to produce image-level decisions. 

Though these methods are effective in classifying histopathological images, the context-capturing ability of saliency-based methods is limited to words and is not able to encode spatial relationships between words. Also, some of these methods (e.g., \cite{hou2016patch,gecer2018detection}) require manual threshold selection to identify  salient regions. The latter segmentation-based methods address these limitations; however, acquiring tissue-level segmentation labels at a large scale is difficult, because experts are required for annotating images. In this work, we introduce a transformer-based method, \arch, to address the limitations of existing methods. Similar to previous work, \arch~is based on the bag-of-words model. However, unlike existing methods, it hierarchically aggregates information at different levels of the model using self-attention, which allows learning of spatial relationships between words and bags. \arch~outperforms existing methods (saliency-based or segmentation-based or their combination) by a significant margin (Section \ref{ssec:main_results}). Moreover, this network pays attention to clinically relevant structures (Section \ref{sec:discussion}).

\vspace{1mm}
\noindent \textbf{Spatial attention:} The most widely studied attention mechanism in visual recognition tasks (image classification, segmentation and object detection) is the spatial attention mechanism \cite{zhou2016learning,selvaraju2017grad}, which weighs the activation maps (or spatial planes) to identify regions of interest. Initially introduced to provide explanations for CNN outputs, variants of this mechanism (supervised \cite{yang2019guided,yao2020saliency} and unsupervised \cite{hu2018squeeze,xu2018structured}) have been incorporated in CNNs to improve the performance across different visual recognition tasks (e.g., \cite{howard2019searching,woo2018cbam}), including medical imaging (e.g., \cite{oktay2018attention,abraham2019novel,schlemper2019attention,rundo2019use,tomita2019attention}). In general, these networks introduce a spatial attention module (e.g., Attention U-Net \cite{oktay2018attention}) within a CNN. Identifying salient regions in histopathological images using spatial attention is difficult because of their large size (usually of the order of gigapixels). This paper introduces an end-to-end transformer-based network for classifying histopathological images.

\vspace{1mm}
\noindent \textbf{Transformers:} Recent work has extended 
transformers \cite{vaswani2017attention} (described in Section \ref{sec:transformer}) for natural images (e.g., classification \cite{parmar2018image}, segmentation \cite{huang2019ccnet,wang2020axial} and object detection \cite{carion2020end}). Though these approaches are effective in learning local and global representations, extending these approaches to histopathological images is challenging primarily because of their large size (e.g., images in our dataset are $45\times$ larger than the ImageNet dataset \cite{russakovsky2015imagenet}). In this work, we extend transformers using bag-of-words models to classify breast biopsy images in an end-to-end fashion. Specifically, we introduce a bottom-up decoding method that allows us to hierarchically refine the information from words to bags to image and produce diagnostic categories. This hierarchical refinement allows clinically relevant tissue structures to be identified without any explicit supervision. To the best of our knowledge, this is the first work that 1) uses transformers for histopathological image analysis and 2) provides explanations for diagnostic decisions.
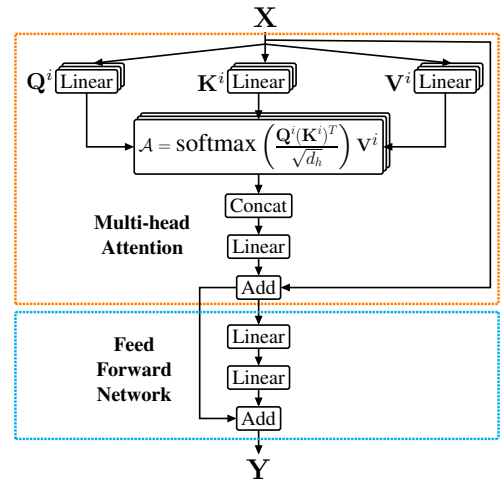
\begin{figure}[b!]
    \centering
    \resizebox{0.75\columnwidth}{!}{
        \newcommand\lw{1mm}
\tikzset{>=triangle 45}

\definecolor{mygreen}{RGB}{51,160,44}

\newcommand{\transformer}{
    \begin{tikzpicture}[block/.style={
inner sep=8,outer sep=0,
align=center,rounded corners, line width=\lw,
font=\fontsize{35}{60}\selectfont}]

    \node [block, fill=white] (x0) at (0, 2.5) {\scalebox{2}{$\mathbf{X}$}};
    \node (x) at (0, 1) {};
    \draw[thick, -, line width=\lw] (x0.south) -- (x.south);

    \node (q3) [block, draw, fill=white] at (0, -1.0) {Linear};
    \node (q2) [block, draw, fill=white] at (-0.2, -1.2) {Linear};
    \node (q1) [block, draw, fill=white, label=left: {\scalebox{4}{$\mathbf{K}^i$}}] at (-0.4, -1.4) {Linear}; %, label=below right: {\huge ($n$, $d$, $d_h$)}
    
    % values
    \node (v3) [block, draw, fill=white] at (6.9+4.5, -1.0) {Linear};
    \node (v2) [block, draw, fill=white] at (6.7+4.5, -1.2) {Linear};
    \node (v1) [block, draw, fill=white, label=left: {\scalebox{4}{ $\mathbf{V}^i$}}] at (6.5+4.5, -1.4) {Linear}; %, label=below right: {\huge ($n$, $d$, $d_h$)}
    
    % keys
    \node (k3) [block, draw, fill=white] at (-6.5-4, -1.0) {Linear};
    \node (k2) [block, draw, fill=white] at (-6.7-4, -1.2) {Linear};
    \node (k1) [block, draw, fill=white, label=left: {\scalebox{4}{$\mathbf{Q}^i$}}] at (-6.9-4, -1.4) {Linear}; %, label=below right: {\huge ($n$, $d$, $d_h$)}
    
    % draw connections between input and K, Q, V
    \draw[thick, ->, line width=\lw] (x.south) -- (q3.north);
    \draw[thick, ->, line width=\lw] (x.south) -- (k3.north);
    \draw[thick, ->, line width=\lw] (x.south) -- (v3.north);

    % draw matrix multiplication 
    \node (kqv3) [block, draw, fill=white] at (0.0, -5) {\scalebox{1.25}{$\mathcal{A} = \text{softmax}\left(\frac{\mathbf{Q}^i (\mathbf{K}^i)^T}{\sqrt{d_h}}\right)\mathbf{V}^i$}};
    \node (kqv2) [block, draw, fill=white] at (-0.2, -5.2) {\scalebox{1.25}{$\mathcal{A} = \text{softmax}\left(\frac{\mathbf{Q}^i (\mathbf{K}^i)^T}{\sqrt{d_h}}\right)\mathbf{V}^i$}};
    \node (kqv1) [block, draw, fill=white] at (-0.4, -5.4) {\scalebox{1.25}{$\mathcal{A} = \text{softmax}\left(\frac{\mathbf{Q}^i (\mathbf{K}^i)^T}{\sqrt{d_h}}\right)\mathbf{V}^i$}};
    
    \draw[thick, ->, line width=\lw] (k1) |- (kqv1);
    \draw[thick, ->, line width=\lw] (q1) -- (kqv1);
    \draw[thick, ->, line width=\lw] (v1) |- (kqv1);
    
    % concatenation
    \node (con) [block, draw] at (-0.4, -9) {Concat};
    \draw[thick, ->, line width=\lw] (kqv1) -- (con);
    
    % Linear
    \node (proj) [block, draw, fill=white] at (-0.4, -11.5) {Linear}; %, label=below right: {\huge ($n$, $d$, $d$)}
    \draw[thick, ->, line width=\lw] (con) -- (proj);
    
    % Residual connection
    \node (add) [block, draw, fill=white] at (-0.4, -14) {Add};
    \draw[thick, ->, line width=\lw] (proj) -- (add);
    \draw[thick, ->, line width=\lw] (x.north) -- (12.75+1, 1) -- (12.75+1, -14) -- (add.east);
    
    % draw MHA BOX
    \draw[dotted, color=orange, thick, line width=1.5*\lw] (-10.75-4.5, 1.5) -- (9.25+5, 1.5);
    \draw[dotted, color=orange, thick, line width=1.5*\lw] (9.25+5, 1.5) -- (9.25+5, -15);
    \draw[dotted, color=orange, thick, line width=1.5*\lw] (-10.75-4.5, -15) -- (9.25+5, -15);
    \draw[dotted, color=orange, thick, line width=1.5*\lw] (-10.75-4.5, -15) -- (-10.75-4.5, 1.5);
    \node (mha) [block] at (-7.5, -10) {{\fontsize{35}{60}\selectfont\textbf{Multi-head}}};
    \node [block, below=0.1cm of mha] {{\fontsize{35}{60}\selectfont\textbf{Attention}}};

    % Draw FFN
    \node (ffn1) [block, draw, fill=white] at (-0.4, -17) {Linear}; %, label=below right: {\huge ($n$, $d$, $d_{ffn}$)}
    \node (ffn2) [block, draw, fill=white] at (-0.4, -19.5) {Linear}; %, label=below right: {\huge ($n$, $d_{ffn}$, $d$)}
    \node (add2) [block, draw, fill=white] at (-0.4, -22) {Add};
    
    \node (out) [block, fill=white, below=1.5cm of add2] {\scalebox{2}{$\mathbf{Y}$}};
    
    \draw[thick, ->, line width=\lw] (add) -- (ffn1);
    \draw[thick, ->, line width=\lw] (ffn1) -- (ffn2);
    \draw[thick, ->, line width=\lw] (ffn2) -- (add2);
    
    % draw residual
    \draw[thick, ->, line width=\lw] (add) -- (-4, -14) -- (-4, -22) -- (add2);
    \draw[thick, ->, line width=\lw] (add2) -- (out);
    
    % draw FFN box
    \draw[dotted, color=cyan, thick, line width=1.5*\lw] (-10.75-4.5, -15.5) -- (9.25+5, -15.5);
    \draw[dotted, color=cyan, thick, line width=1.5*\lw] (9.25+5, -15.5) -- (9.25+5, -23.25);
    \draw[dotted, color=cyan, thick, line width=1.5*\lw] (-10.75-4.5, -23.25) -- (9.25+5, -23.25);
    \draw[dotted, color=cyan, thick, line width=1.5*\lw] (-10.75-4.5, -23.25) -- (-10.75-4.5, -15.5);
    \node (mha) [block] at (-8, -19) {{\fontsize{35}{60}\selectfont\textbf{Forward}}};
    \node (mha1) [block, below=0.1cm of mha] {{\fontsize{35}{60}\selectfont\textbf{Network}}};
    \node (mha2) [block, above=0.1cm of mha] {{\fontsize{35}{60}\selectfont\textbf{Feed}}};
    
    \end{tikzpicture}
}\transformer
    }
    \caption{The Transformer unit stacks multi-head attention and a feed forward network to model interactions between inputs.}
    \label{fig:transformer_unit}
\end{figure}
\begin{figure*}[t!]
    \centering
    \resizebox{2\columnwidth}{!}{
        \centering
	    \input{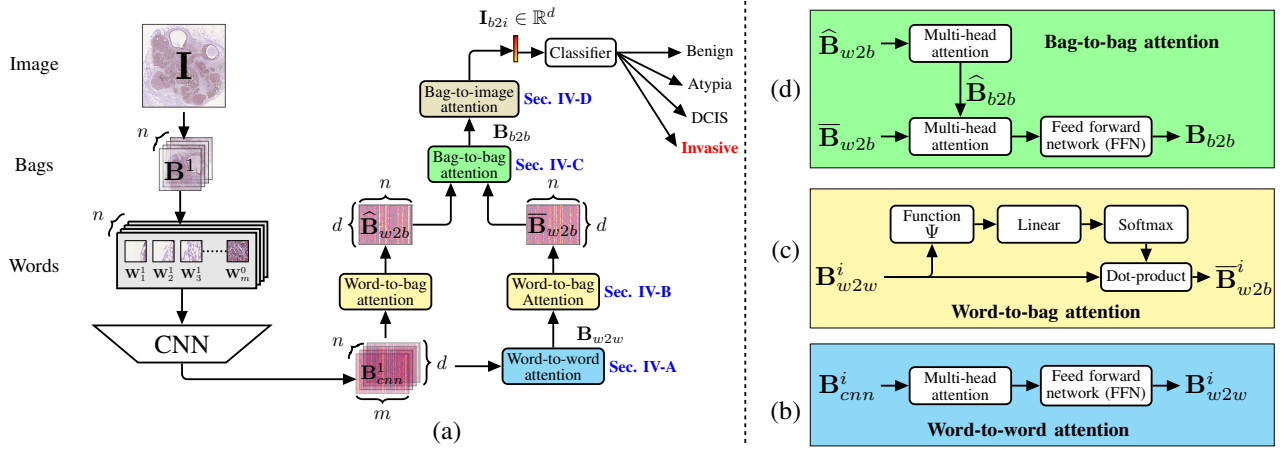}\archiectureNew
    }
    \caption{(a) \arch: Our end-to-end holistic attention network for classifying breast biopsy whole slide images models the relationships between bags and words in a hierarchical manner using self attention. (b-d) Word-to-word, word-to-bag, and bag-to-bag attention modules are visualized that allows learning relationships between bags and words using a bottom-up method, respectively. Note that word-to-bag attention module for processing $\mathbf{B}_{cnn}$ and bag-to-image attention module for processing $\mathbf{B}_{b2b}$ are similar to (c) and therefore, we do not visualize here.}
    \label{fig:hollistic_attn_net}
\end{figure*}
\section{Background: Transformers}
\label{sec:transformer}
Transformer-based attention networks \cite{vaswani2017attention} allow inputs to interact with each other, so that the model can automatically find important inputs on which to focus. The transformer module, shown in Figure \ref{fig:transformer_unit}, consists of two parts: (1) multi-head attention that models relationships between inputs, and (2) a feed forward network that learns wider representations. 

The multi-head attention module has three projection branches that maps the input $\mathbf{X} \in \mathbb{R}^{N \times d}$ to query ($\mathbf{Q}$), key ($\mathbf{K}$), and value ($\mathbf{V}$), where $N$ is the number of inputs (words and bags in this case) and $d$ is the input dimensionality. In particular, each branch consists of $H$ linear layers (also called heads) that projects $\mathbf{X}$ from $d$- to $d_h$-dimensional space, allowing us to learn multiple views of the input, where $d_h = d/H$. To allow the module to attend over different input positions, it computes $H$ dot-products between keys ($\mathbf{K}$) and queries ($\mathbf{Q}$) and produces $H$ attention weight matrices, each of size $N \times N$. These resultant $H$ attention weight matrices encode the relationships between $N$ inputs and are combined with the values ($\mathbf{V}$) after a softmax using another dot-product to produce weighted sum outputs. The independent $H$ outputs are then concatenated and fused using another linear layer $\boldsymbol \beta_{mha} \in \mathbb{R}^{d \times d}$ to produce the output $\mathbf{X}_{mha} \in \mathbb{R}^{N \times d}$. The multi-head attention operation is defined as:
\begin{equation}
    \begin{array}{rl}
         \mathbf{X}_{mha} &= \text{MultiHead}(\mathbf{X}_Q=\mathbf{X},  \mathbf{X}_K=\mathbf{X}, \mathbf{X}_V=\mathbf{X}) \\ 
          & = \text{Concat}(head_1, \cdots, head_H) \boldsymbol \beta_{mha}
    \end{array}
    \label{eq:mha}
\end{equation}
where $head_i=\mathcal{A}( \underbrace{\mathbf{X}_Q \boldsymbol \beta_Q^i}_{\mathbf{Q}^i},\ \underbrace{\mathbf{X}_K \boldsymbol \beta_K^i}_{\mathbf{K}^i},\ \underbrace{\mathbf{X}_V \boldsymbol \beta_V^i}_{\mathbf{V}^i})$ is the output of the $i$-th head and $\mathcal{A} = \text{softmax}\left(\frac{\mathbf{Q}^i (\mathbf{K}^i)^T}{\sqrt{d_h}}\right) \mathbf{V}^i$ is the scalar dot-product attention. 

The multi-head attention module learns narrower representations in $d_h$-dimensional space. To enable the transformer unit to learn wider representations, the output of multi-head attention $\mathbf{X}_{mha}$ is fed to the feed forward network (FFN). The FFN is a stack of two linear layers. The first linear layer with weights $\boldsymbol \beta_E \in \mathbb{R}^{d \times 4d}$ expands the input from $d$- to $4d$-dimensional space, while the second linear layer with weights $\boldsymbol \beta_R \in \mathbb{R}^{4d \times d}$ projects back from $4d$- to $d$-dimensional space. Mathematically, we can define this operation as:
\begin{equation}
    \mathbf{Y} = \text{FFN}(\mathbf{X}_{mha}) = \text{ReLU}(\mathbf{X}_{mha}\ \boldsymbol \beta_E) \boldsymbol \beta_R
\end{equation}
This work extends the transformer unit using the bag-of-word model to encode inter-word and inter-bag relationships. This increases the context-capturing ability of the network and improves performance.

\section{HATNet: Holistic Attention Network}
\label{sec:method}
State-of-the-art CNN-based classification networks (e.g., ResNet \cite{he2016deep}) stack convolutional layers and down-sampling layers to learn representations at multiple scales. These networks are difficult to apply to histopathological images, primarily because the resolution of these medical images (e.g., 10K $\times$ 10K) are much larger than images used in standard image classification tasks (e.g., $224 \times 224$). To address this resolution challenge, a standard approach is to learn patch-wise (or word-wise) representations using a sliding window method (e.g., \cite{hou2016patch,mehta2018ynet,gecer2018detection}). Though these approaches have shown to be effective for histopathological image classification, the context-capturing ability of such approaches is still limited to patch-level. Also, such approaches are difficult to train in an end-to-end manner. 

This paper introduces an end-to-end approach that allows diagnostic class prediction using the entire histopathological image at once. Our method extends the transformer architecture \cite{vaswani2017attention} using a bag-of-words approach and is shown in Figure \ref{fig:hollistic_attn_net}. We call our model a Holistic ATtention Network (\arch) because of its ability to learn inter-word and inter-bag representations in an end-to-end fashion. With {\it attention}, we emphasize the progressive hierarchical refining from words to bags to image to produce the classification output.

Briefly, \arch~first encodes inter-word representations using self-attention (Section \ref{ssec:w2w_attn}). These representations are then combined to produce bag-level representations (Section \ref{ssec:w2b_attn}). We then encode inter-bag representations (Section \ref{ssec:b2b_attn}), which are then combined to produce image-level representations (Section \ref{ssec:b2i_attn}). These representations are classified to produce the diagnosis category (Section \ref{ssec:diag_class}). Because of the bottom-up decoding (words $\rightarrow$ bags $\rightarrow$ image), representations learned using \arch~are expressive and allow the identification of important words and bags in an image. We believe that this will help us build tools to annotate clinically important words and explain  diagnosis decisions. 

\begin{figure*}[t!]
    \centering
    \begin{subfigure}[b]{\columnwidth}
        \centering
        \includegraphics[width=0.9\columnwidth]{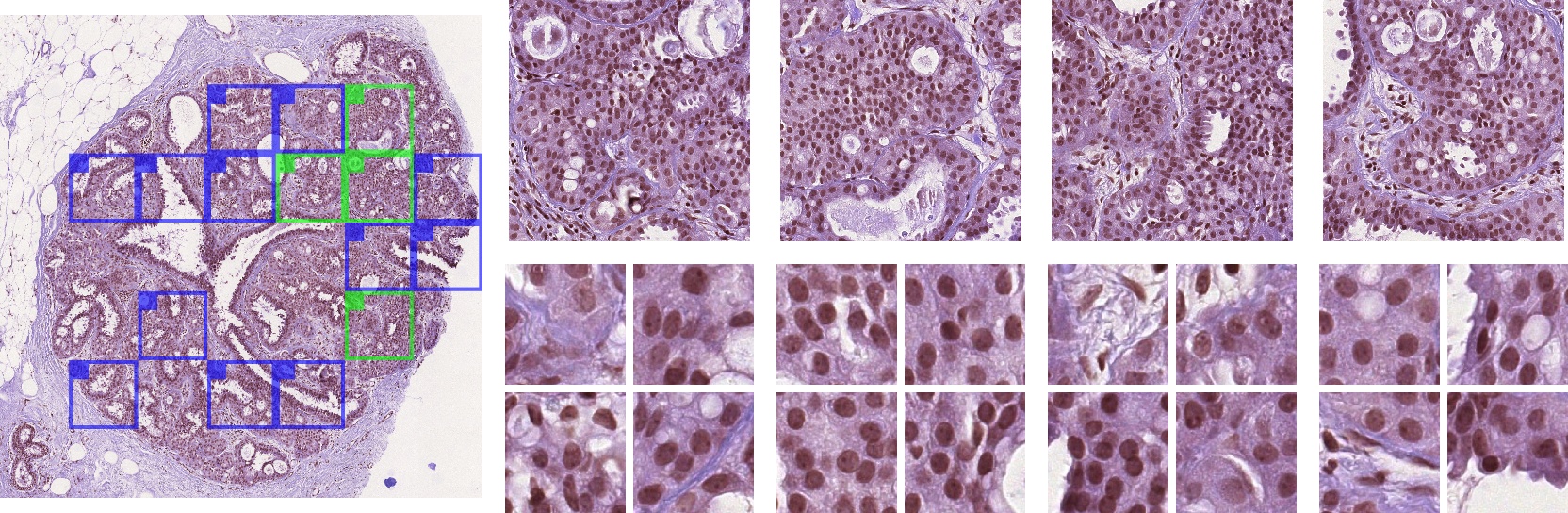} 
        \caption{Benign}
    \end{subfigure}
    \hfill
    \begin{subfigure}[b]{\columnwidth}
        \centering
        \includegraphics[width=0.9\columnwidth]{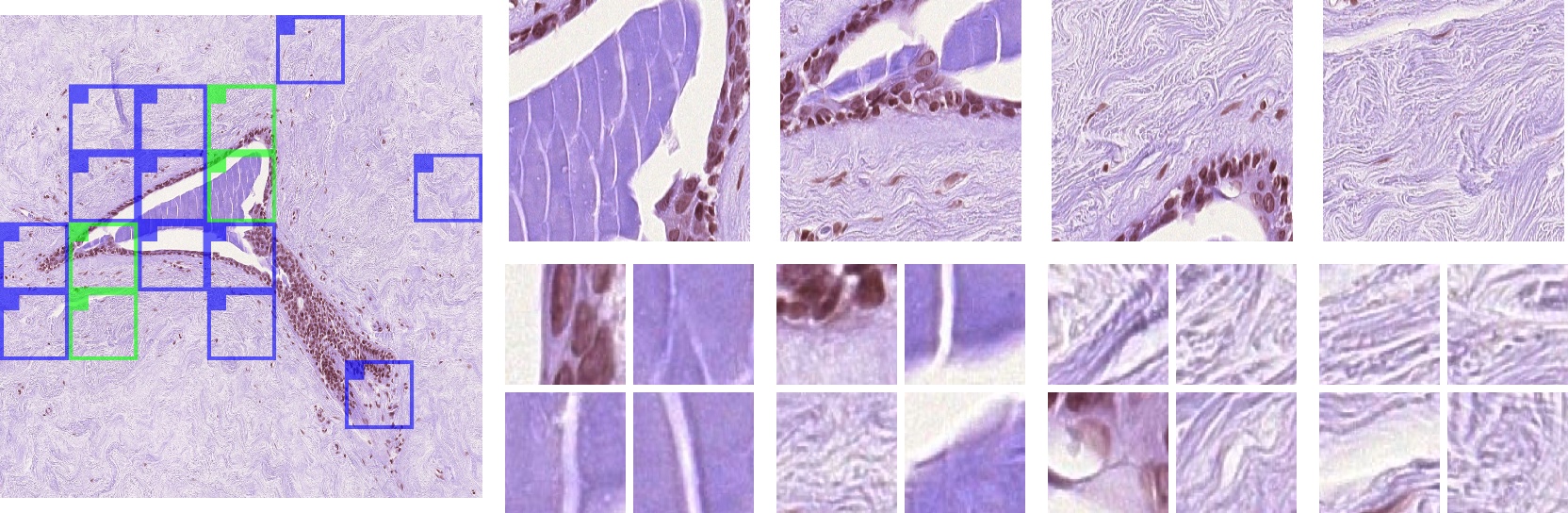} 
        \caption{Atypia}
    \end{subfigure}
    \vfill
    \begin{subfigure}[b]{\columnwidth}
        \centering
        \includegraphics[width=0.9\columnwidth]{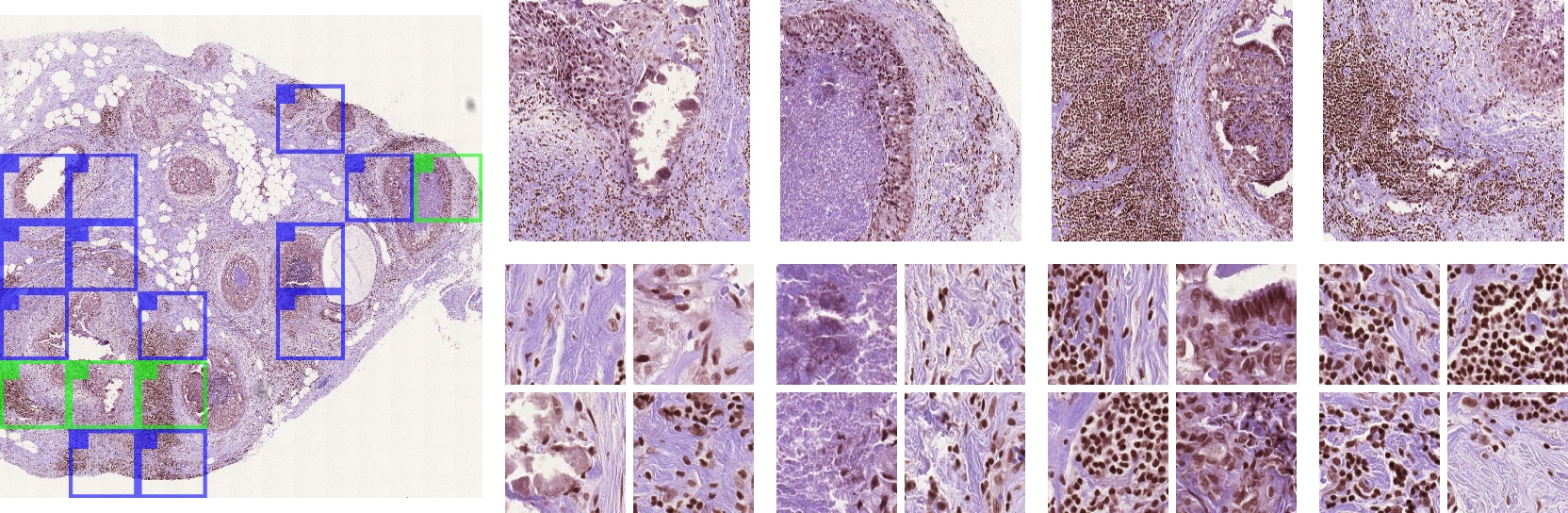} 
        \caption{Ductal carcinoma in-situ (DCIS)}
    \end{subfigure}
    \hfill
    \begin{subfigure}[b]{\columnwidth}
        \centering
        \includegraphics[width=0.9\columnwidth]{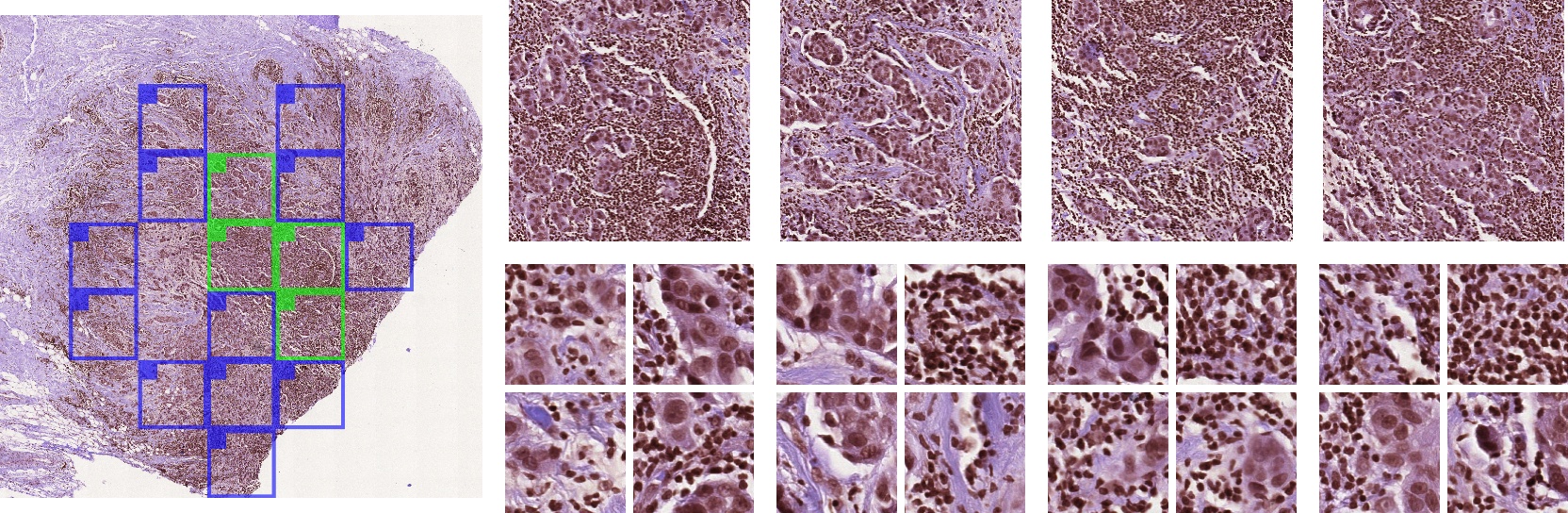} 
        \caption{Invasive}
    \end{subfigure}
    \caption{Example results of bags and words identified using \arch~across different diagnostic categories. \arch~aggregates information from different parts of the image and different textures. Here, each sub-figure of the breast biopsy image is shown on the left of each panel with the top-30\% bags (top-4 in \textcolor{applegreen}{\bf green}, the rest in \textcolor{blue}{\bf blue}) identified using \arch~overlayed on image. The upper right in each panel shows the top-4 bags, and the bottom right in each panel shows the top-4 words in each bag.} 
    \label{fig:bag_word_examples}
\end{figure*}

\subsection{Word-to-word attention}
\label{ssec:w2w_attn}
% first high level idea; then what you did; then the conclusion and what it leads to
The word-to-word attention module (see Figure \ref{fig:hollistic_attn_net}(b)) is comprised of a transformer unit (Section \ref{sec:transformer}) with multi-head attention and a feed-forward network, allowing us to model the interactions between words and identify important words in the whole slide image. 

The input image $\mathbf{I} \in \mathbb{R}^{w \times h}$ with width $w$ and height $h$ is first divided into $n$ non-overlapping  bags $\mathbf{I} = \left(\mathbf{B}^1, \cdots, \mathbf{B}^n \right) \in \mathbb{R}^{\frac{w}{n} \times \frac{h}{n}}$, where $\mathbf{B}^i$ represents the $i$-th bag. Each bag $\mathbf{B}^i$ is then divided into $m$ non-overlapping  words $\mathbf{B}^i = \left( \mathbf{W}_1^i, \cdots, \mathbf{W}_m^i \right) \in \mathbb{R}^{\frac{w}{nm} \times \frac{h}{nm}}$, where $\mathbf{W}_j^i$ represents the $j$-th word in the $i$-th bag. Similar to previous works (e.g. \cite{hou2016patch,mehta2018ynet}), we feed words $\mathbf{W}_j^i$ inside each bag $\mathbf{B}^i$ to a CNN to produce word-level representations for each bag: $\mathbf{B}_{cnn}^i = (\mathbf{\widehat{W}}_1^i, \cdots, \mathbf{\widehat{W}}_m^i ) \in \mathbb{R}^{d}$. The representations from the CNN does not encode inter-word relationships. We encode inter-word relationships in each bag $\mathbf{B}_{cnn}^i$ using the transformer unit (Section \ref{sec:transformer}) to produce $\mathbf{B}_{w2w}^i \in \mathbb{R}^{m \times d}$ as:
\begin{equation}
    \resizebox{0.91\columnwidth}{!}{
    $\mathbf{B}_{w2w}^i = \text{FFN}\left( \text{Multihead}(\mathbf{X}_Q=\mathbf{B}_{cnn}^i, \mathbf{X}_K=\mathbf{B}_{cnn}^i, \mathbf{X}_V=\mathbf{B}_{cnn}^i) \right)$
    }
\end{equation}
The key-query-value decomposition in multi-head attention allows us to encode inter-word relationships, and the FFN allows us to learn wider representations.

\subsection{Word-to-bag attention}
\label{ssec:w2b_attn}
The word-to-word attention identifies relevant words in each bag. We aggregate these word-level representations to produce bag-level representations (see Figure \ref{fig:hollistic_attn_net}(c)). To do so, we linearly combine the words inside each bag $\mathbf{B}_{w2w}^i$. In particular, we first map each word in $\mathbf{B}_{w2w}^i$ from $\mathbb{R}^{d}$ to $\mathbb{R}^{1}$ using a projection function $\Psi$. Since each bag has $m$ words, this projection function $\Psi$ produces a vector of length $m$. We then apply a linear transformation $\overline{\boldsymbol \beta}_{w2b} \in \mathbb{R}^{m \times m}$ and softmax to produce $m$ coefficients, which are then used to linearly combine words in  $\mathbf{B}_{w2w}^i$ to produce bag-level representation $\mathbf{\overline{B}}_{w2b}^i \in \mathbb{R}^{d}$ as:
\begin{equation}
    \mathbf{\overline{B}}_{w2b}^i = \text{softmax}\left( \Psi(\mathbf{B}_{w2w}^i)\   \overline{\boldsymbol \beta}_{w2b} \right) \mathbf{B}_{w2w}^i, \quad 1 \le i \le n
    \label{eq:w2b_sa}
\end{equation}

Similarly, we can combine word-level representations obtained from the CNN for each bag $\mathbf{B}_{cnn}^i$ using $\Psi$ and linear transformation $\widehat{\boldsymbol \beta}_{w2b} \in \mathbb{R}^{m \times m}$ to produce bag-level representations, $\mathbf{\widehat{B}}_{w2b}^i \in \mathbb{R}^{d}$. 
\begin{equation}
    \mathbf{\widehat{B}}_{w2b}^i = \text{softmax}\left( \Psi( \mathbf{B}_{cnn}^i)\  \widehat{\boldsymbol \beta}_{w2b} \right) \mathbf{B}_{cnn}^i, \quad 1 \le i \le n
     \label{eq:w2b_cnn}
\end{equation}

\subsection{Bag-to-bag attention}
\label{ssec:b2b_attn}
The bag-level representations $\mathbf{\overline{B}}_{w2b}$ and $\mathbf{\widehat{B}}_{w2b}$ do not encode information about surrounding bags. To encode inter-bag relationships, we apply bag-to-bag attention (see Figure \ref{fig:hollistic_attn_net}(d)). The bag-to-bag attention module is similar to the word-to-word attention (Section \ref{ssec:w2w_attn}), except that we use $\mathbf{\widehat{B}}_{w2b}$ (Eq. \ref{eq:w2b_cnn}) as context to $\mathbf{\overline{B}}_{w2b}$ (Eq. \ref{eq:w2b_sa}). This also mimics the typical skip-connection mechanism in neural networks \cite{he2016deep,ronneberger2015u}.

We first apply multi-head attention to $\mathbf{\widehat{B}}_{w2b}$ to encode inter-bag representations and produce $\mathbf{\widehat{B}}_{b2b} \in \mathbb{R}^{n \times d}$ as: 
\begin{equation}
    \resizebox{0.91\columnwidth}{!}{
        $\mathbf{\widehat{B}}_{b2b} = \text{Multihead}(\mathbf{X}_Q=\mathbf{\widehat{B}}_{w2b},  \mathbf{X}_K=\mathbf{\widehat{B}}_{w2b}, \mathbf{X}_V= \mathbf{\widehat{B}}_{w2b})$
    }
\end{equation}
To allow every bag obtained from CNN $\mathbf{\widehat{B}}_{b2b}$ to attend over every bag $\mathbf{\overline{B}}_{w2b}$ obtained after word-level self-attention, we apply another multi-head attention in which $\mathbf{\widehat{B}}_{b2b}$ serves as a query and  $\mathbf{\overline{B}}_{w2b}$ serves as keys and values, allowing us to encode rich inter-bag  representations and produce $\mathbf{B}_{b2b} \in \mathbb{R}^{n \times d}$.  Mathematically, we can define the bag-to-bag attention operation as:
\begin{equation}
    \resizebox{0.91\columnwidth}{!}{
    $\mathbf{B}_{b2b} = \text{FFN}\left( \text{Multihead}(\mathbf{X}_Q=\mathbf{\widehat{B}}_{b2b},  \mathbf{X}_K=\mathbf{\overline{B}}_{w2b}, \mathbf{X}_V= \mathbf{\overline{B}}_{w2b}) \right)$
    }
\end{equation}

\subsection{Bag-to-image attention}
\label{ssec:b2i_attn}
The inter-bag representations are encoded in $\mathbf{B}_{b2b} \in \mathbb{R}^{n \times d}$. We aggregate these representations to produce image-level representations. Similar to word-to-bag attention (Section \ref{ssec:w2b_attn}), we combine these bag-level representations using a function $\Psi$ and linear transformation $\boldsymbol \beta_{b2i} \in \mathbb{R}^{n \times n}$ to produce image-level representations $\mathbf{I}_{b2i} \in \mathbb{R}^{d}$.
\begin{equation}
    \mathbf{I}_{b2i} = \text{softmax}\left( \Psi( \mathbf{B}_{b2b} )\  \boldsymbol \beta_{b2i} \right) \mathbf{B}_{b2b}
     \label{eq:b2i_cnn}
\end{equation}
Because of the bottom-up decoding (words to bags to image), these representations are expressive and allows us to identify important words and bags in an image (Figure \ref{fig:bag_word_examples}).

\subsection{Classification and Loss}
\label{ssec:diag_class}
We classify $\mathbf{I}_{b2i}\in \mathbb{R}^d$ into $C$-diagnosis classes using a linear classifier with weights $\boldsymbol \beta_{cls} \in \mathbb{R}^{d \times C}$ as:
\begin{equation}
    \hat{y} = \text{softmax}\left(\mathbf{I}_{b2i} \ \boldsymbol \beta_{cls}\right)
\end{equation}
We minimize the cross-entropy loss $\mathcal{L}$ between the ground truth $y$ and prediction $\hat{y}$ to train \arch. During evaluation, we choose the index that has the highest confidence score in $\hat{y}$ as our predicted class label. %Note that $0 < \hat{y}_i < 1$ and $\sum\limits_{i=1}^C \hat{y}_i = 1$.

\section{Experimental Results}
\label{sec:exp_results}

\subsection{Dataset and evaluation} 
\label{ssec:model_dataset}
The breast biopsy dataset \cite{elmore2015diagnostic} consists of 240 whole slide images with haematoxylin and eosin (H\&E) staining. The image dataset was designed to include a higher prevalence of cases from diagnostic categories that have lower prevalence in the general population. This provides a robust and more challenging image dataset. A total of 87 pathologists participated in a previous study and interpreted these cases. Each pathologist classified a random subset of 60 slides into four diagnostic categories (benign, atypia, ductal carcinoma in-situ (DCIS), and invasive breast cancer). This resulted in 22 diagnostic labels (on average) per slide. A group of three expert pathologists then interpreted these cases and provided a consensus label per slide. We treat these consensus labels as ground truth diagnostic labels. The pathologists also marked 420 regions of interest (ROIs) that best supported the diagnoses. Following previous studies on this dataset \cite{mercan2017multi,mehta2018learning,mehta2018ynet,mercan2019assessment,gecer2018detection}, we use these ROIs to train and evaluate our method using the following metrics: (1) classification (or top-1) accuracy, (2) F1-score, (3) sensitivity, (4) specificity, and (5) area under receiver operating characteristic curve (ROC-AUC). In particular, the dataset consists of 164 training, 42 validation, and 216 test ROIs (see  Table \ref{tab:data_splits}). %We use the following metrics to assess classification performance: 

\begin{table}[b!]
    \centering
    \resizebox{0.85\columnwidth}{!}{
        \begin{tabular}{lccccc}
            \toprule[1.5pt]
            \textbf{Diagnostic} & \multicolumn{4}{c}{\textbf{Number of ROIs}} & \textbf{Average size} \\
            \cmidrule[1.25pt]{2-5}
            \textbf{Category} & \textbf{Training} & \textbf{Validation} & \textbf{Test} & \textbf{Total} & \textbf{(in pixels)}\\
            \midrule[1pt]
            Benign & 48 & 13 & 64 & 125 & 6731 $\times$ 5839 \\
            Atypia & 40 & 8 & 54 & 102 & 10668 $\times$ 8967 \\
            DCIS & 60 & 17 & 84 & 161 & 10778 $\times$ 9547 \\
            Invasive & 16 & 4 & 14 & 34 & 23866 $\times$ 21402 \\
            \midrule[1pt]
            Total & 164 & 42 & 216 & 422 & 10880 $\times$ 9558 \\
            \bottomrule[1.5pt]
        \end{tabular}
    }
    \caption{Statistics of breast biopsy whole slide image dataset.}
    \label{tab:data_splits}
\end{table}

\subsection{Architecture}
\label{ssec:model_arch}
The size of the ROIs is variable. To use computational resources efficiently, each ROI is resized to  spatial dimensions of about $12544 \times 12544$. The resized ROIs are then split into $n=49$ non-overlapping bags, each bag with a spatial dimension of $1792 \times 1792$. Each bag is further split into $m=49$ non-overlapping words, resulting in words with a spatial dimension of $256 \times 256$. Overall, each ROI has 2401 words. These words are fed to off-the-shelf CNNs to extract word-level representations. In our experiments, three state-of-the-art light-weight CNNs pretrained on the ImageNet dataset \cite{deng2009imagenet} are studied: (1) ESPNetv2 \cite{mehta2019espnetv2}, (2) MobileNetv2 \cite{sandler2018mobilenetv2}, and (3) MNASNet \cite{tan2019mnasnet}. ESPNetv2 follows an Inception-style design \cite{szegedy2015going} and uses four simultaneous $3\times3$ depth-wise convolutions with different dilation rates. MobileNetv2 follows ResNet-style design \cite{he2016deep}. To improve the computational efficiency, MobileNetv2 uses $3\times3$ depth-wise convolutions instead of $3\times3$ standard convolutions. MNASNet uses the same basic building block as MobileNetv2; however, it uses neural architecture search \cite{zoph2018learning} to identify the optimal model configuration, which provides best trade-off between different parameters. The proposed network is generic and any off-the-shelf heavy-weight CNNs (e.g., ResNet \cite{he2016deep}) can be used to extract word-level representations. Heavy-weight networks are not explored because of computational constraints.

The dimensionality of word-level representations varies from CNN to CNN. Therefore, the output of a CNN is linearly projected to a 256 dimensional space ($d=256$). To encode the inter-word and inter-bag representations, 4 heads ($H=4$) are used in multi-head attention, resulting in a head dimension of $d_h=64$. We use the function $\Psi$ to aggregate word-level representations into bag-level representations (Section \ref{ssec:w2b_attn}) and bag-level representations into image-level representations (Section \ref{ssec:b2i_attn}). In our experiments, we study three different functions: (1) Euclidean distance (or L2 norm), (2) Manhattan distance (or L1 norm), and (3) mean of a vector.

\subsection{Training}
\label{ssec:model_training}
Our models are trained end-to-end using the ADAM \cite{kingma2014adam} optimizer with a learning rate warm-up strategy. We first warm-up the learning rate from $10^{-7}$ to $10^{-4}$ in 600 iterations and then train the model for the next 50 epochs with a learning rate of $10^{-4}$. After that, learning rate is decayed by half and then train for another 50 epochs. Our model takes about 36 hours for training on two NVIDIA GeForce GTX 1080 GPUs, each with a memory of 8 GB. We accumulate gradients for 8 iterations before updating the weights, yielding an effective batch size of 8 ROIs per update. Training data is augmented by randomly resizing ($192\times 192$,  $224\times 224$, $256\times 256$, $288\times 288$, $320\times 320$), flipping, and rotating (angle: $-10^\circ$ to $10^\circ$) the words. For evaluation, a single model is obtained by averaging the best 5 validation checkpoints. Compared to the best model, averaged models delivered 1-1.5 points higher accuracy.
\begin{table*}[t!]
    \centering
    \resizebox{2\columnwidth}{!}{
        \begin{tabular}{llrrrrrrrr}
            \toprule[1.5pt]
             \multicolumn{1}{l}{\textbf{Row}} & \multicolumn{1}{c}{\multirow{2}{*}{\textbf{Model}}} & \multicolumn{2}{c}{\textbf{Parameters}} & \multicolumn{1}{c}{ } & \multicolumn{5}{c}{\textbf{Evaluation metrics}} \\
            \cmidrule[1pt]{3-4} \cmidrule[1pt]{6-10}
             \multicolumn{1}{l}{\textbf{No.}} & & \multicolumn{1}{c}{\textbf{CNN}} & \multicolumn{1}{c}{\textbf{Attn.}} & \multicolumn{1}{c}{\textbf{ }} & \multicolumn{1}{c}{\textbf{Accuracy}} & \multicolumn{1}{c}{\textbf{F1-score}} & \multicolumn{1}{c}{\textbf{Sensitivity}} & \multicolumn{1}{c}{\textbf{Specificity}} & \multicolumn{1}{c}{\textbf{ROC-AUC}} \\
            \midrule[1pt]
            R1 & Pathologists (avg. of 70 pathologists) \cite{mehta2018ynet} & \multicolumn{2}{c}{ } && 0.70 & 0.71 & 0.70 & 0.90  &  \\
            \midrule
            R2 & LAB \& LBP hand-crafted features (w/o saliency)  & \multicolumn{2}{r}{ }  && 0.28 &  &  &  &  \\
            R3 &  LAB \& LBP hand-crafted features (w/ saliency)  & \multicolumn{2}{r}{ }  && 0.45 &  &  &  &  \\
            \midrule
            R4 &  Bag-of-word (majority voting w/o saliency)  & \multicolumn{2}{r}{ } && 0.23 &  &  &  &  \\
            R5 & Bag-of-word (majority voting w/ saliency)  & \multicolumn{2}{r}{ }  && 0.55 &  &  &  &  \\
            R6 & Bag-of-word (learned fusion w/o saliency)  & \multicolumn{2}{r}{ }  && 0.38 &  &  &  &  \\
            R7 & Bag-of-word (learned fusion w/ saliency)  & \multicolumn{2}{r}{ }  && 0.55 &  &  &  &  \\
            \midrule
            R8 & MRSegNet with histogram and co-occurence features  & 26.03 M & NA && 0.55 & 0.56 & 0.55 &  0.85 &  \\
            R9 & MRSegNet with structural features & 26.03 M & NA && 0.56 & 0.57 & 0.56 & 0.85 &  \\
            R10 & Y-Net & 3.91 M & NA && 0.62 & 0.62 & 0.62 & 0.87 &  \\
            \midrule
            R11 & \arch~(w/ ESPNetv2) & 2.21 M & 2.37 M && 0.67 & 0.64 & 0.67 & 0.89 & 0.89 \\
            R12 & \arch~(w/ MobileNetv2) & 2.22 M & 2.37 M && 0.66 & 0.65 & 0.66 & 0.89 & 0.88 \\
            R13 & \arch~(w/ MNASNet) & 3.10 M & 2.37 M && \textbf{0.70} & \textbf{0.70} & \textbf{0.70} & \textbf{0.90} & \textbf{0.90} \\
            \midrule
            R14 & \arch~(Ensemble) & NA & NA && \textbf{0.71} & \textbf{0.70} & \textbf{0.71} & \textbf{0.90} & \textbf{0.90} \\
            \bottomrule[1.5pt]
        \end{tabular}
    }
    \caption{\textbf{Comparison with state-of-the-art networks.} \arch~outperforms existing methods by a significant margin. Network parameters are reported for single models only. We use majority voting for ensembling the models.}
    \label{tab:comparison_sota}
\end{table*}

\subsection{Baseline networks}
\label{ssec:baseline_networks}
We compare our method with the following methods:
\begin{enumerate}
    \item \textbf{Bag-of-words model with hand-crafted features \cite{gecer2018detection}:}  An input image (bag) is split into words. Following \cite{basavanhally2013multi}, LAB histogram and LBP histogram features are extracted from these words. These word-level features are concatenated and then classified using logistic regression into diagnostic categories with and without saliency (see R2 and R3 in Table \ref{tab:comparison_sota}). Similar to a standard practice in saliency-based approaches (e.g.,  \cite{hou2016patch,wang2018weakly,mehta2018ynet}), the class with majority voting in saliency maps is selected as a diagnostic category.
    \item \textbf{Bag-of-words model with deep features \cite{gecer2018detection}:} Instead of using hand-crafted features, this approach uses a deep convolutional neural network, FCN \cite{long2015fully}, to obtain word-level representations. These representations are used to identify discriminative or salient regions. In addition to majority voting-based method, a learned fusion method \cite{hou2016patch} is also tried to model the relationships between words (see R4-R7 in Table \ref{tab:comparison_sota}).
    \item \textbf{Multi-resolution segmentation network (MRSegNet) \cite{mehta2018learning}:} MRSegNet has two stages: (1) tissue-level segmentation and (2) diagnostic classification. The first stage is a multi-resolution encoder-decoder network which combines the outputs of many words (or patches) at different resolutions to reduce segmentation errors. In the second stage, histogram and co-occurrence features are extracted from tissue-level segmentation masks, which are then classified using a multi-layer perceptron into different diagnostic classes (see R8 in Table \ref{tab:comparison_sota}). 
    \item \textbf{Structural features \cite{mercan2019assessment}:} This method extracts structural features from tissue-level segmentation masks produced using MRSegNet. These features allows capturing structural changes in ductal regions, an important biomarker for breast cancer diagnosis \cite{zhang2012guidelines,kinne1989breast,page1998ductal,shah2016management}, in the breast (see R9 in Table \ref{tab:comparison_sota}).
    \item \textbf{Y-Net \cite{mehta2018ynet}:} Y-Net adds a classification branch to U-Net \cite{ronneberger2015u} and allows to jointly predict the tissue-level segmentation mask and the saliecy map. The saliency map is then combined with the tissue-level segmentation mask to produce a discriminative segmentation mask. Similar to MRSegNet, Y-Net extracts histogram and co-occurrence features, which are then used for classify diagnostic classes (see R10 in Table \ref{tab:comparison_sota}).
\end{enumerate}

\subsection{Main Results}
\label{ssec:main_results}
Table \ref{tab:comparison_sota} shows that \arch~outperforms state-of-the-art methods significantly and delivers performance comparable to pathologists. For example, \arch~(R13) improves the performance of the best saliency-based models (R5, R7) by about 15\%. When compared to approaches that uses tissue-level segmentation masks (R8-R10) to capture the structural changes in biopsies, \arch~delivers better performance. In particular, \arch~improves the F1-score of the previous best segmentation-based approach (R10) by 8\%. Overall, these results shows that \arch~is effective. We note that ensembling these three models (R14) further improves the accuracy and senstivity by 1\%. 

Furthermore, Table \ref{tab:inference_time} shows that \arch~is fast. \arch~with MNASNet is about $1.8\times$ faster than the previously best reported network, i.e., Y-Net.
\begin{table}[t!]
    \centering
    \resizebox{0.85\columnwidth}{!}{
        \begin{tabular}{lrr}
            \toprule[1.5pt]
             \textbf{Model} & \textbf{Accuracy} & \multicolumn{1}{c}{\textbf{Inference time}} \\
             \midrule[1.25pt]
             Y-Net & 0.62 & $3.93 \text{ s} \pm 20 \text{ ms}$\\
             \arch~(w/ ESPNetv2) & 0.67 & $2.63 \text{ s} \pm 19 \text{ ms}$\\
             \arch~(w/ MobileNetv2)  & 0.66 & $2.17 \text{ s} \pm 10 \text{ ms}$ \\
             \arch~(w/ MNASNet)  & \textbf{0.70} & $\mathbf{2.13 \text{ s} \pm 12 \text{ ms}}$\\
             \bottomrule[1.5pt]
        \end{tabular}
    }
    \caption{\textbf{Inference time}. \arch~is fast and accurate compared to previous best model (Y-Net). Inference time is measured on a single NVIDIA GTX 1080 Ti GPU and is an average across 100 trails on the validation set. }
    \label{tab:inference_time}
\end{table}
\begin{figure*}[t!]
    \centering
    \begin{subfigure}[b]{0.66\columnwidth}
        \centering
        \includegraphics[width=0.83\columnwidth]{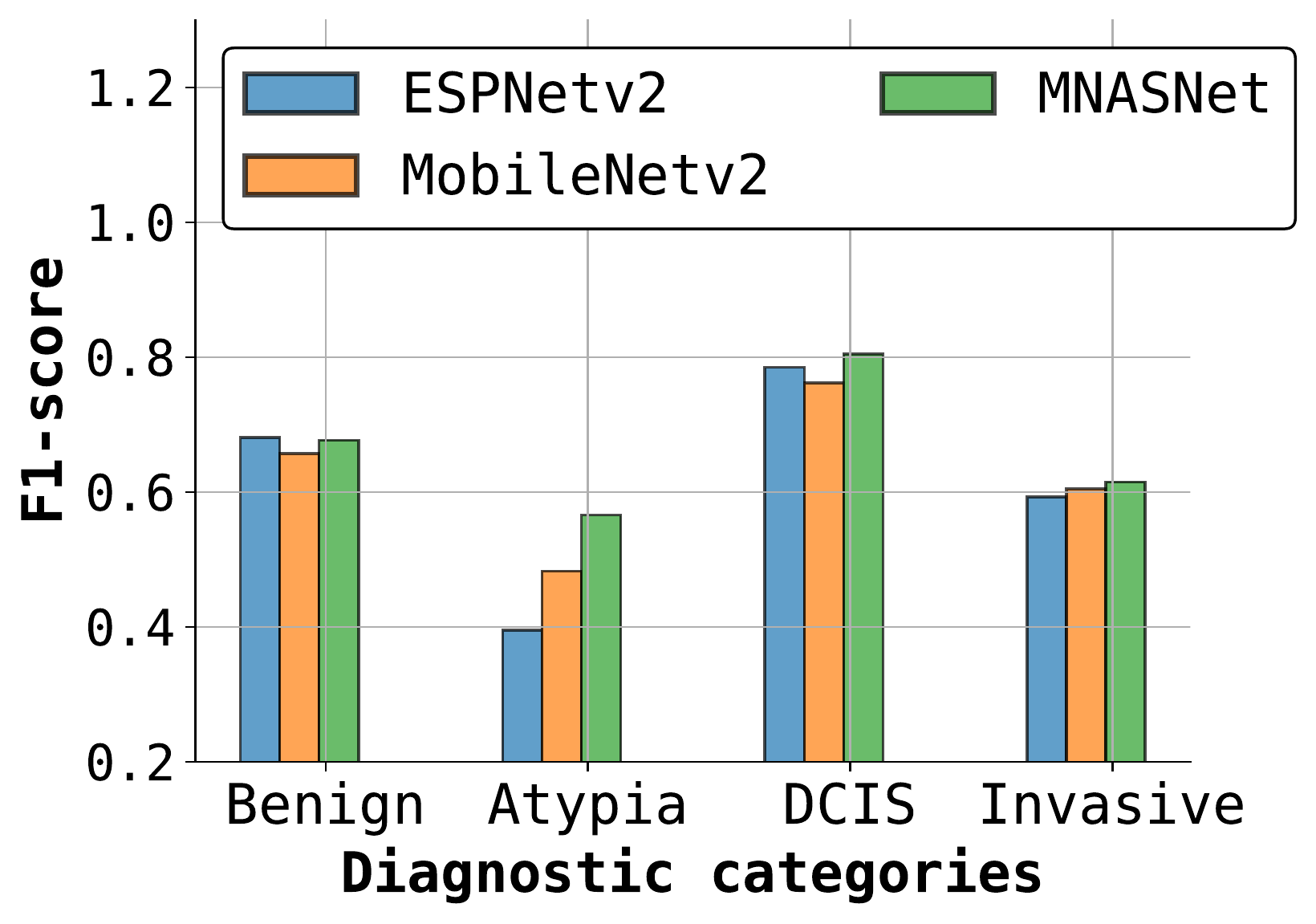}
        \caption{F1-score}
        \label{fig:f1_class_wise}
    \end{subfigure}
    \hfill
    \begin{subfigure}[b]{0.66\columnwidth}
        \centering
        \includegraphics[width=0.83\columnwidth]{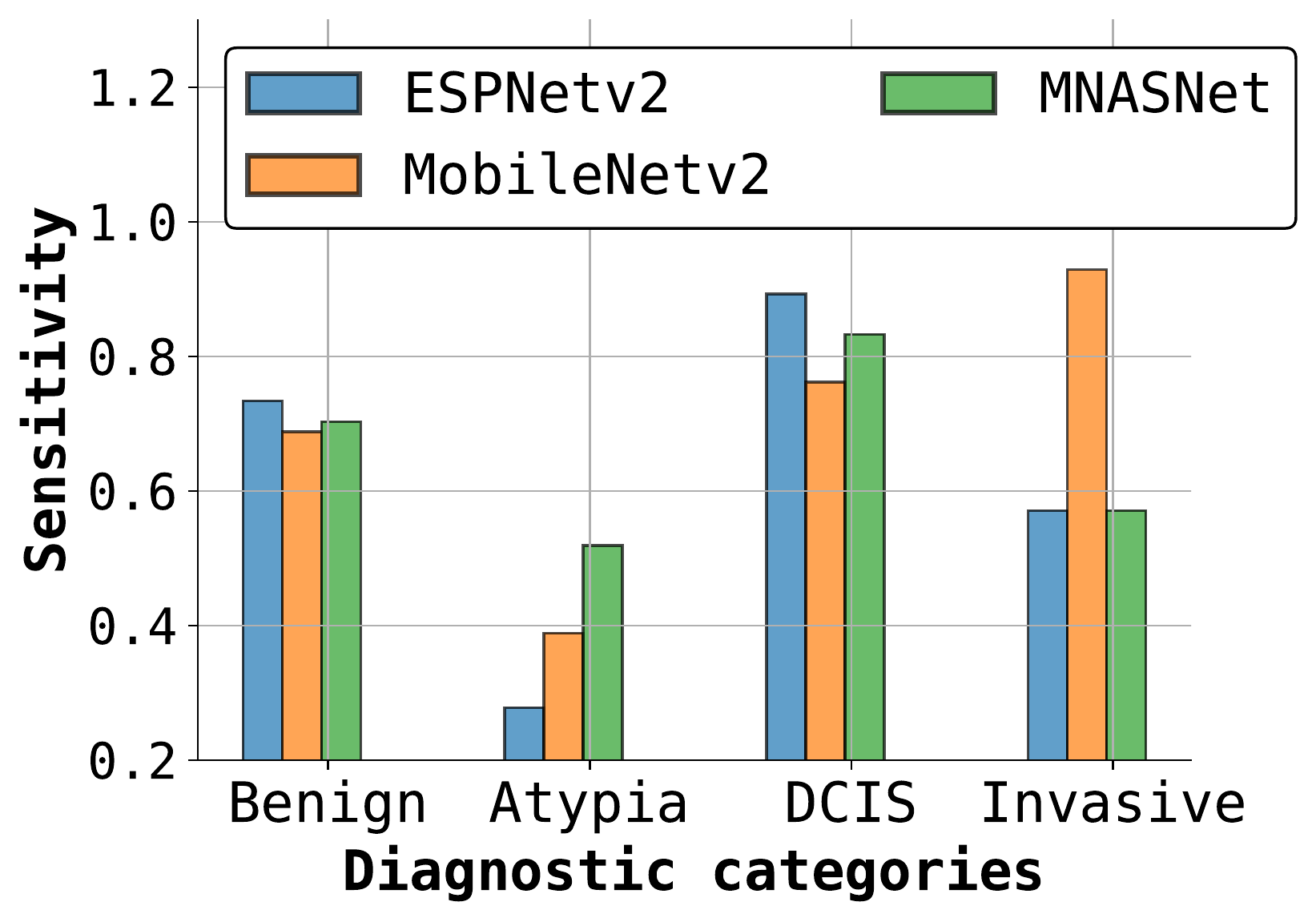}
        \caption{Sensitivity}
        \label{fig:sens_class_wise}
    \end{subfigure}
    \hfill
    \begin{subfigure}[b]{0.66\columnwidth}
        \centering
        \includegraphics[width=0.83\columnwidth]{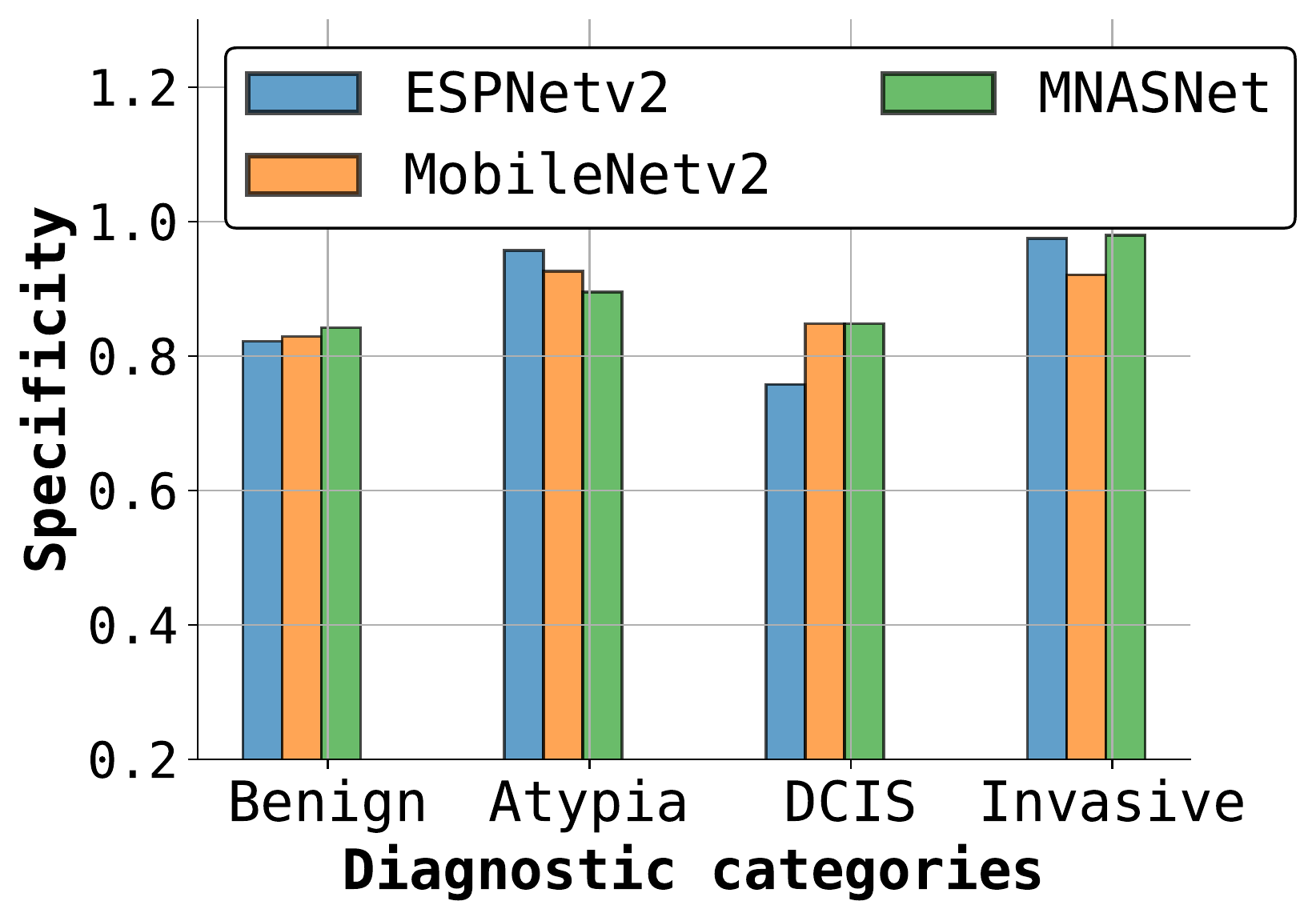}
        \caption{Specificity}
        \label{fig:spec_class_wise}
    \end{subfigure}
    \caption{Class-wise performance comparison of \arch~with different CNN architectures. Overall, the models with MNASNet as a base feature extractor performs a little better than the other two networks across different metrics. However, MNASNet has a low sensitivity score for Invasive Cancer, while MobileNetv2 does much better in this regard.}
    \label{fig:class_wise_metrics}
\end{figure*}
\begin{figure*}[t!]
    \centering
    \begin{subfigure}[b]{0.66\columnwidth}
        \centering
        \includegraphics[width=\columnwidth]{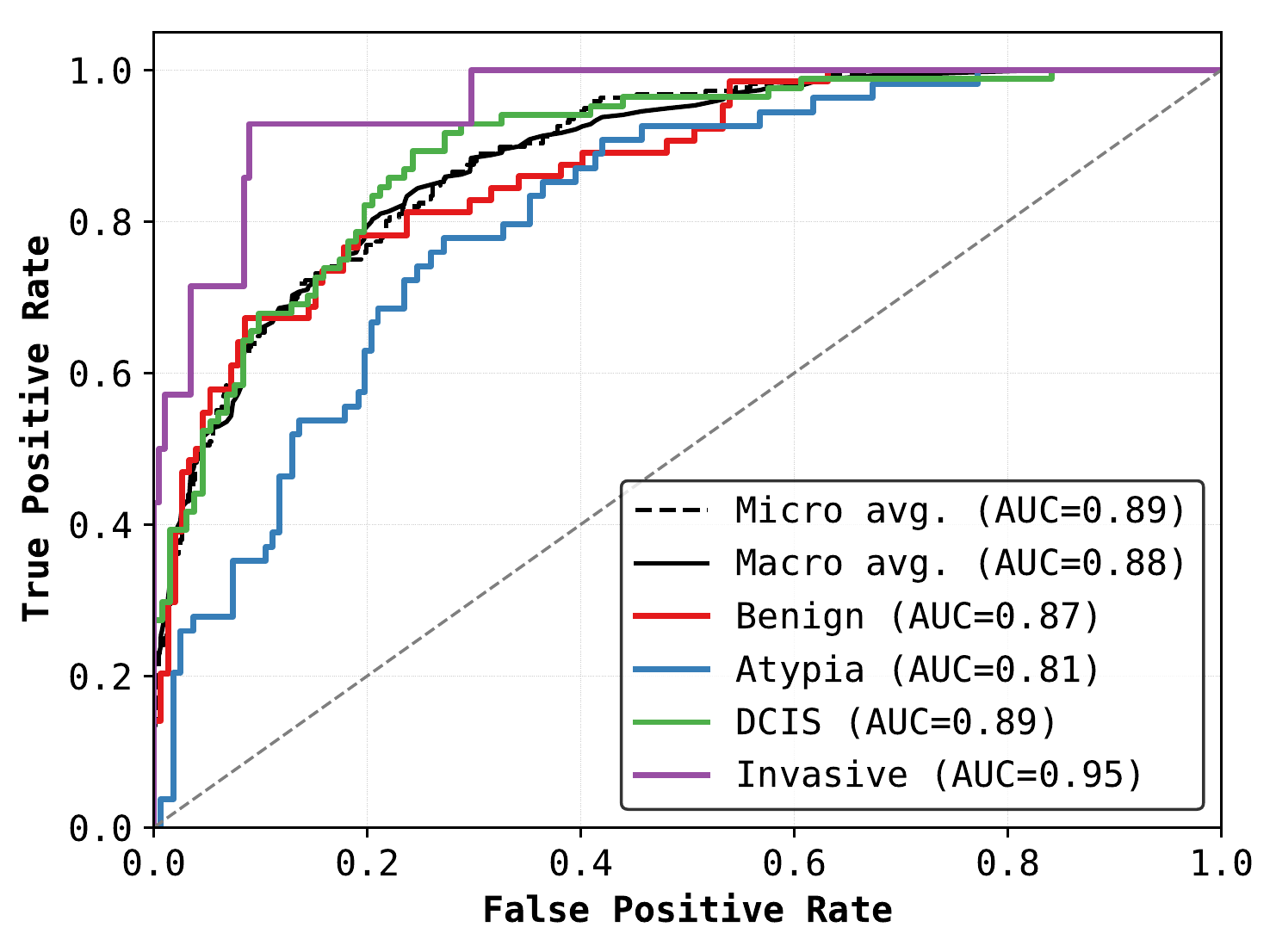}
        \caption{ESPNetv2}
        \label{fig:roc_curve_espnetv2}
    \end{subfigure}
    \hfill
    \begin{subfigure}[b]{0.66\columnwidth}
        \centering
        \includegraphics[width=\columnwidth]{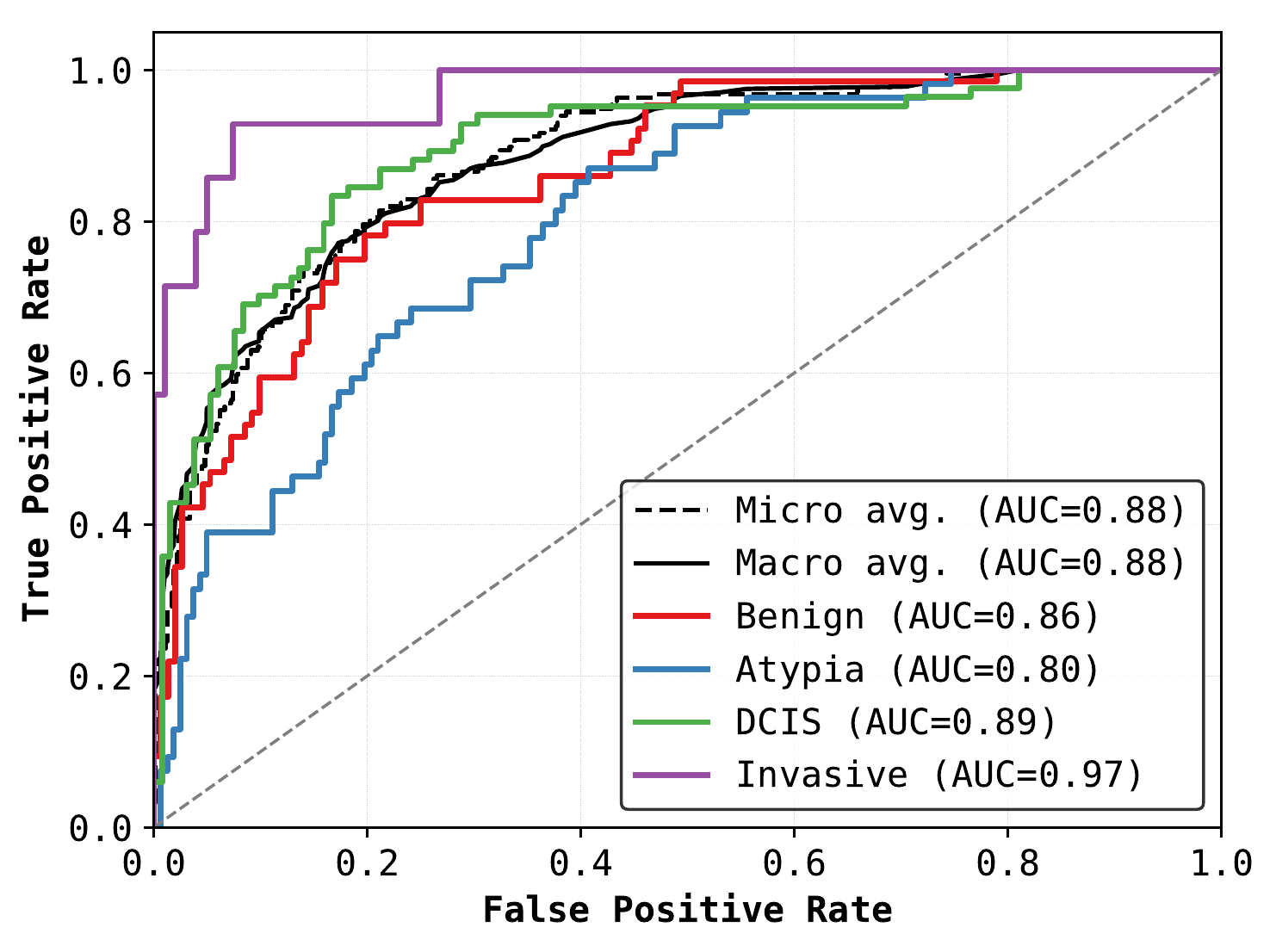}
        \caption{MobileNetv2}
        \label{fig:roc_curve_mobilenetv2}
    \end{subfigure}
    \hfill
    \begin{subfigure}[b]{0.66\columnwidth}
        \centering
        \includegraphics[width=\columnwidth]{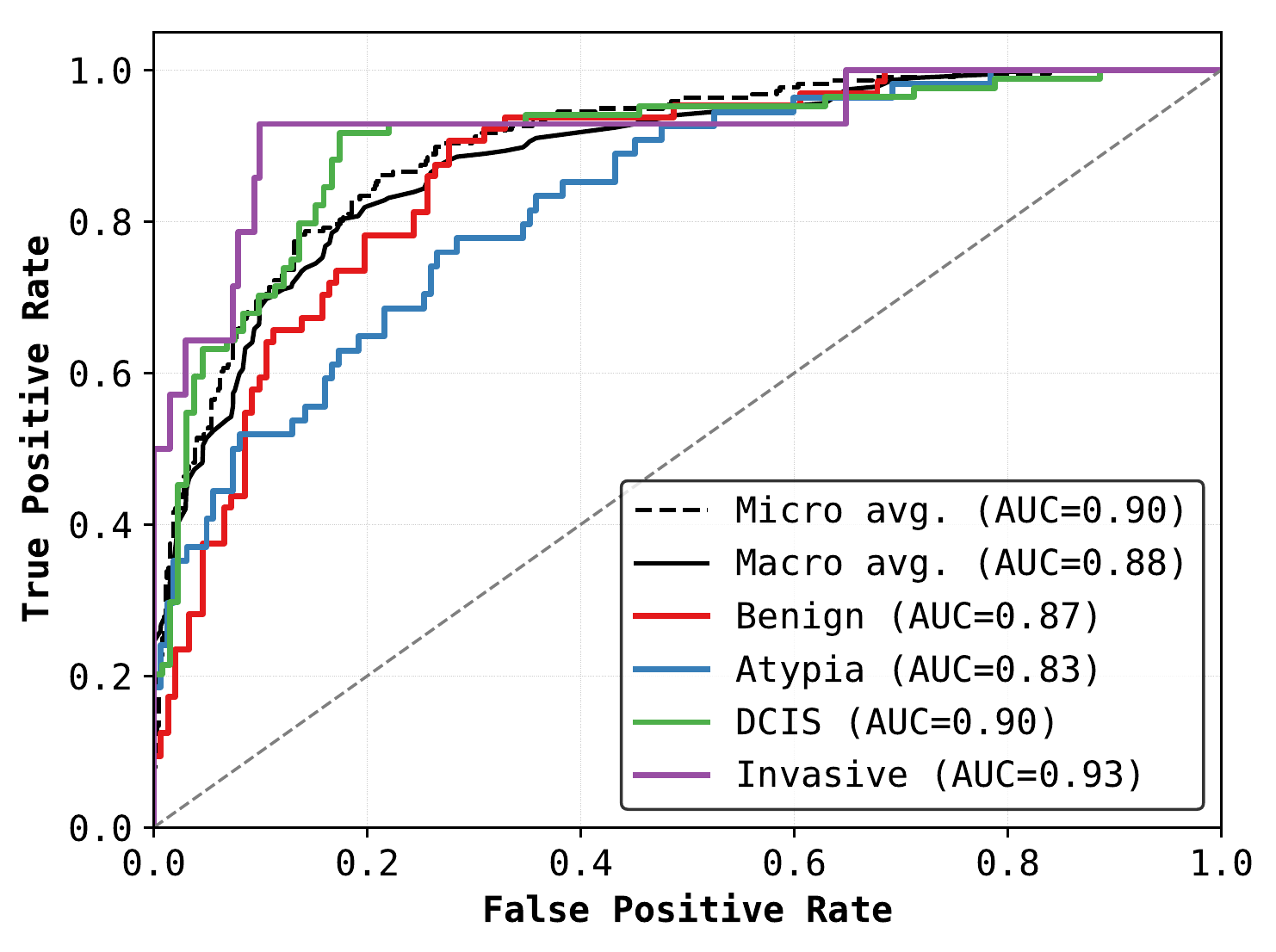}
        \caption{MNASNet}
        \label{fig:roc_curve_mnasnet}
    \end{subfigure}
    \caption{Receiver operating characteristic (ROC) curves of \arch~with different CNN architectures. The models with MNASNet as a base feature extractor has slightly higher area under curve (AUC) than the other two.}
    \label{fig:roc_curves}
\end{figure*}
\subsection{Model Ablations}
\label{ssec:ablations}
\label{ssec:model_results}
\noindent \textbf{Effect of function $\Psi$:} Table \ref{tab:compare_psi} compares the performance of three different $\Psi$ functions with ESPNetv2 as a base feature extractor. Euclidean distance delivers the best performance. The model has 1\% higher accuracy, sensitivity, and specificity values compared to the other two functions. In the rest of the experiments, we use Euclidean distance as a $\Psi$ function.

\vspace{1mm}
\noindent \textbf{Effect of bag and word sizes:} Table \ref{tab:bag_word_sizes} compares the performance of our model with three different bag-word size configurations using ESPNetv2 as a base feature extractor. The bag size of 1792 and word size of 256 delivered slightly better performance than the others. In the rest of the experiments, we use this bag-word size configuration.
\begin{table}[t!]
    \centering
    \resizebox{\columnwidth}{!}{
    \begin{tabular}{lrrrrr}
        \toprule[1.5pt]
       \textbf{Function } $\Psi$  & \textbf{Accuracy} & \textbf{F1-score} & \textbf{Senstivity} & \textbf{Specificity} & \textbf{ROC-AUC} \\
       \midrule[1pt]
       Euclidean distance & \underline{0.67} & 0.64 & \underline{0.67} & \underline{0.89} & 0.89 \\
       Manhattan distance & 0.66 & 0.64 & 0.66 & 0.88 & 0.89 \\
       Mean               & 0.66 & 0.64 & 0.66 & 0.88 & 0.89 \\
       \bottomrule[1.5pt]
    \end{tabular}
    }
    \caption{Effect of different $\Psi$ functions.}
    \label{tab:compare_psi}
\end{table}
\begin{table}[t!]
    \centering
    \resizebox{\columnwidth}{!}{
        \begin{tabular}{ccrrrrr}
            \toprule[1.5pt]
             \textbf{Bag size} & \textbf{Word size} & \textbf{Accuracy} &  \textbf{F1-score} & \textbf{Senstivity} & \textbf{Specificity} & \textbf{ROC-AUC} \\
            \midrule[1pt]
            $1792 \times 1792$  & $256 \times 256$  & 0.67 & 0.64 & 0.67& 0.89& 0.89\\
            $2016 \times 2016$ & $288 \times 288$  & 0.67 & 0.64&  0.67& 0.89 &0.88\\
            $2240 \times 2240$ & $320 \times 320$  & 0.66 & 0.64 & 0.66 & 0.88 & 0.89\\
            \bottomrule[1.5pt]
        \end{tabular}
    }
    \caption{Effect of different bag and word sizes. Note the number of words in each configuration are the same (i.e. 49).}
    \label{tab:bag_word_sizes}
\end{table}

\vspace{1mm}
\noindent \textbf{Effect of different base feature extractors:} Figure \ref{fig:class_wise_metrics} compares the class-wise performance of \arch~with three different base feature extractors. \arch~with MNASNet delivers similar or better class-wise F1-score, sensitivity, and specificity values, except for the invasive case where MobileNetv2 has a higher sensitivity value.

Figure \ref{fig:roc_curves} plots the overall and class-wise receiver operating characteristics of our model with different base feature extractors. We observe that MNASNet delivers the best performance (higher ROC-AUC) compared to the other two networks. Similarly, in Table \ref{tab:comparison_sota} (R11-R13), our model delivers the best overall performance with MNASNet across different evaluation metrics. For example, our model with MNASNet has 6\% and 5\% higher F1-score than with ESPNetv2 and MobileNetv2, respectively.

\begin{figure*}[t!]
    \centering
    \begin{subfigure}[b]{0.45\columnwidth}
        \centering
        \includegraphics[height=65px, width=\columnwidth]{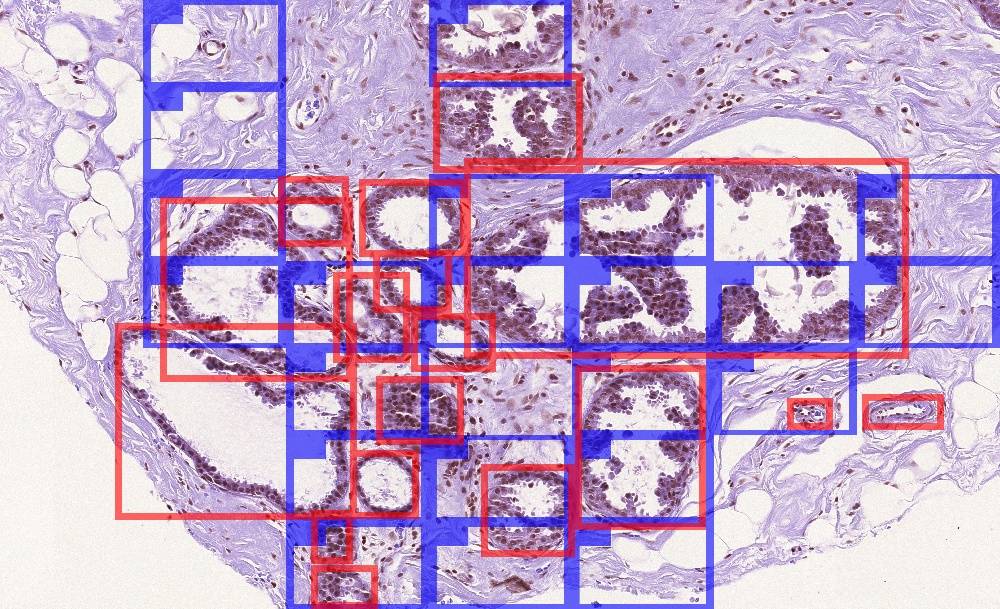}
        \caption{}
        \label{fig:duct_analysis_a}
    \end{subfigure}
    \hfill
    \begin{subfigure}[b]{0.45\columnwidth}
        \centering
        \includegraphics[height=65px, width=\columnwidth]{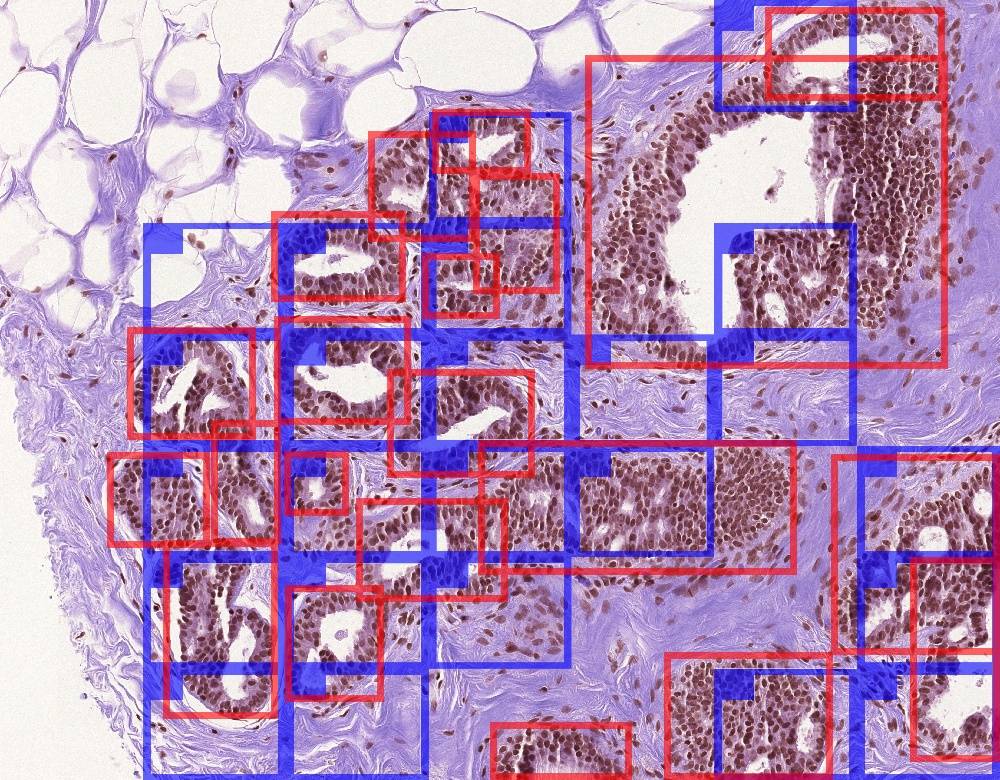}
        \caption{}
        \label{fig:duct_analysis_b}
    \end{subfigure}
    \hfill
    \begin{subfigure}[b]{0.45\columnwidth}
        \centering
        \includegraphics[height=65px, width=\columnwidth]{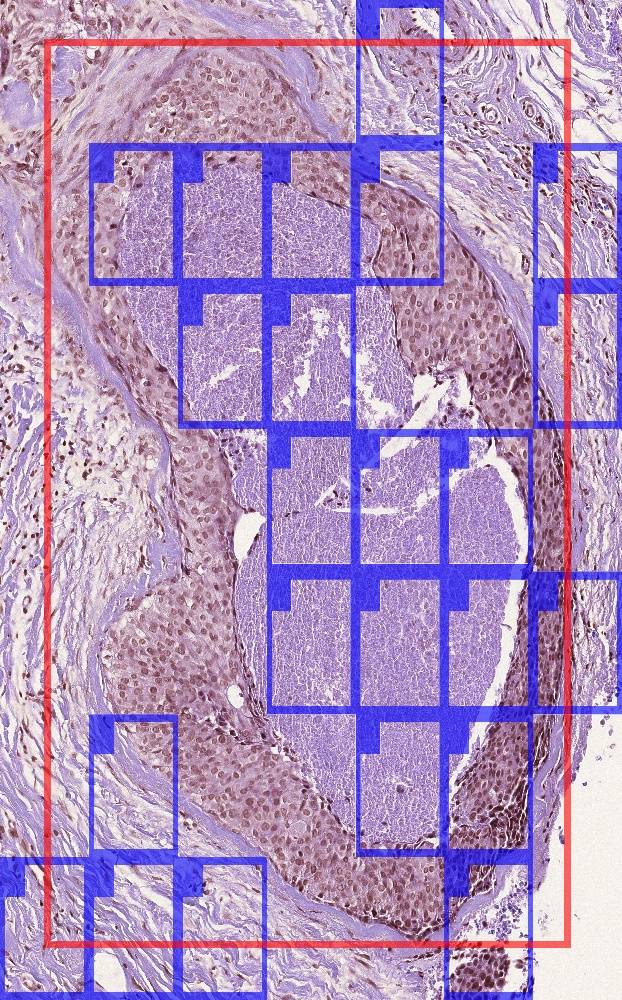}
        \caption{}
        \label{fig:duct_analysis_c}
    \end{subfigure}
    \hfill
    \begin{subfigure}[b]{0.45\columnwidth}
        \centering
        \includegraphics[height=65px, width=\columnwidth]{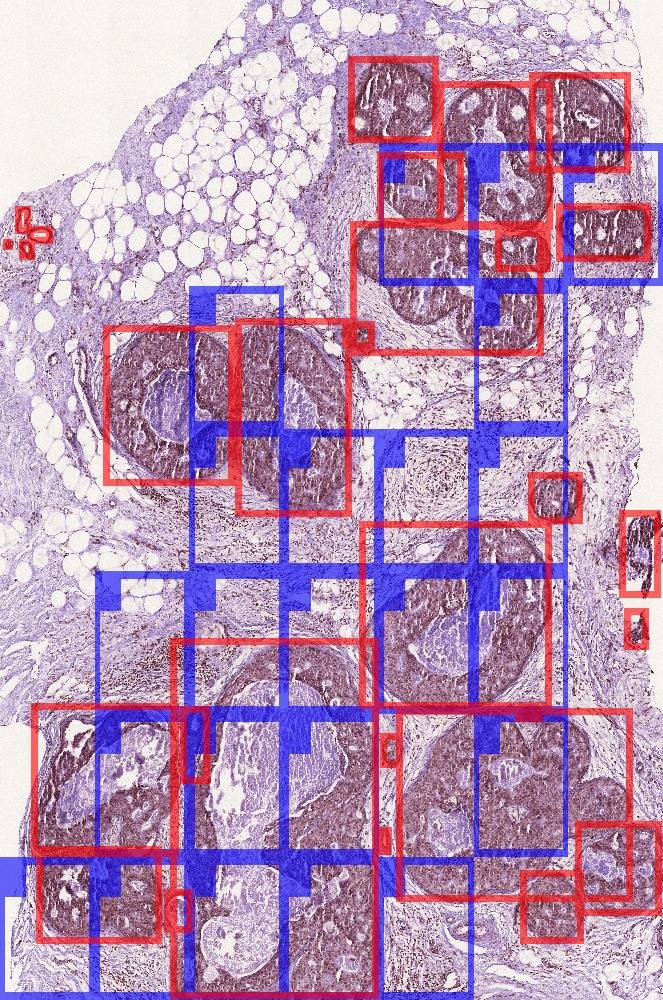}
        \caption{}
        \label{fig:duct_analysis_d}
    \end{subfigure}
    \vfill
    \begin{subfigure}[b]{0.5\columnwidth}
        \centering
        \includegraphics[height=75px, width=\columnwidth]{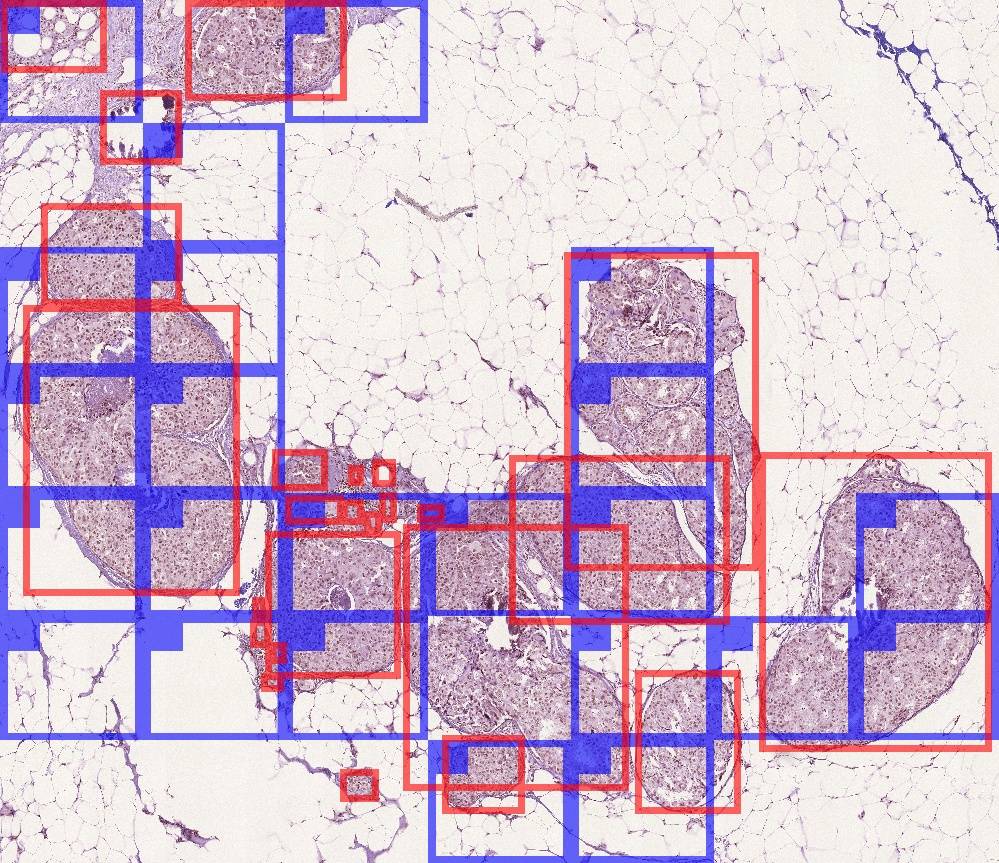}
        \caption{}
        \label{fig:duct_analysis_e}
    \end{subfigure}
    \hfill
    \begin{subfigure}[b]{0.5\columnwidth}
        \centering
        \includegraphics[height=75px, width=\columnwidth]{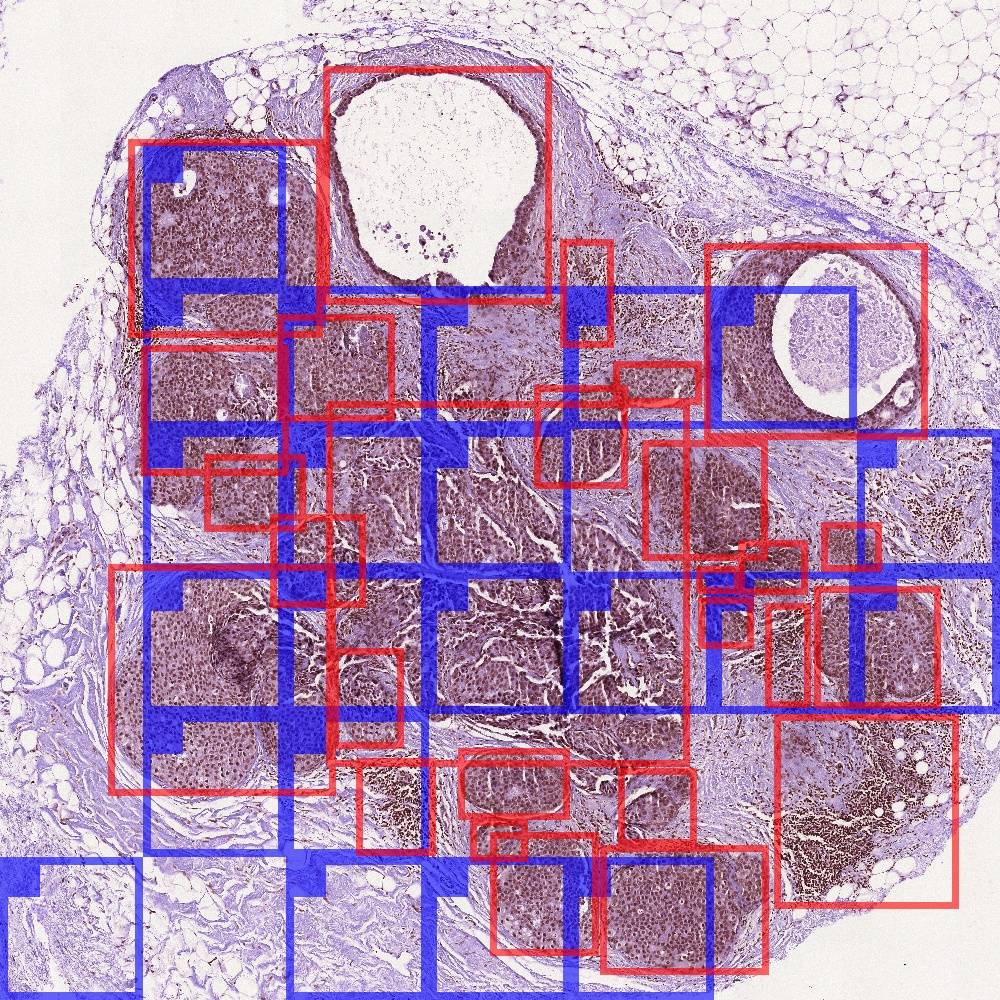}
        \caption{}
        \label{fig:duct_analysis_f}
    \end{subfigure}
    \hfill
    \begin{subfigure}[b]{0.98\columnwidth}
        \centering
        \includegraphics[width=0.7\columnwidth]{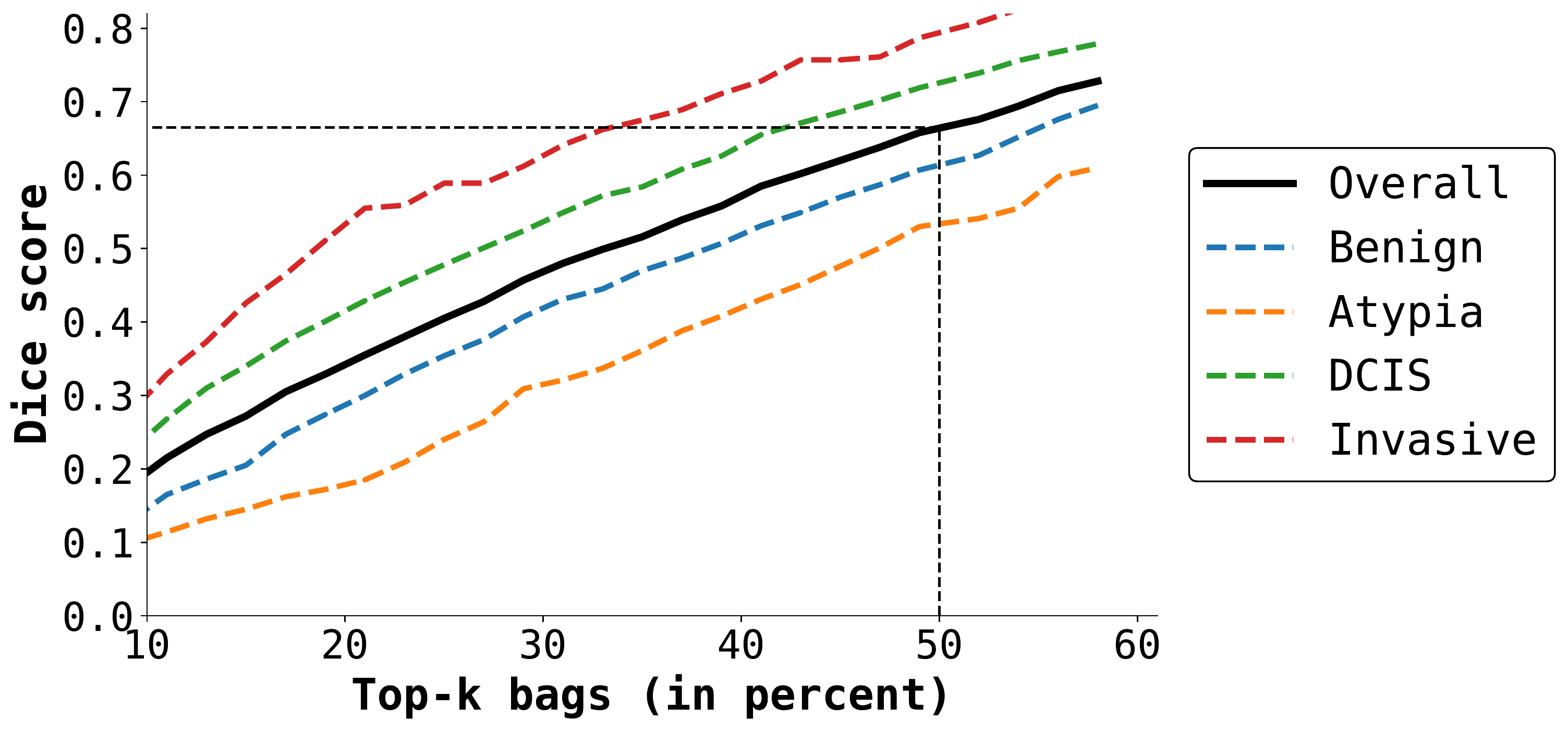}
        \caption{}
        \label{fig:duct_analysis_g}
    \end{subfigure}
    \caption{\arch~identifies ducts of variable size and texture as an important structure. In (a-f), \textcolor{red}{ductal regions} (marked by pathologists) are shown in red, while the \textcolor{blue}{top-50\% bags} predicted by \arch~are shown in blue. In (g), the dice score is plotted between ductal regions and top-k bag predictions (k varies from 10\% to 60\%) for different diagnostic classes.}
    \label{fig:duct_analysis}
\end{figure*}
\begin{figure*}[t!]
    \centering
    \begin{subfigure}[b]{0.4\columnwidth}
        \centering
        \includegraphics[height=70px]{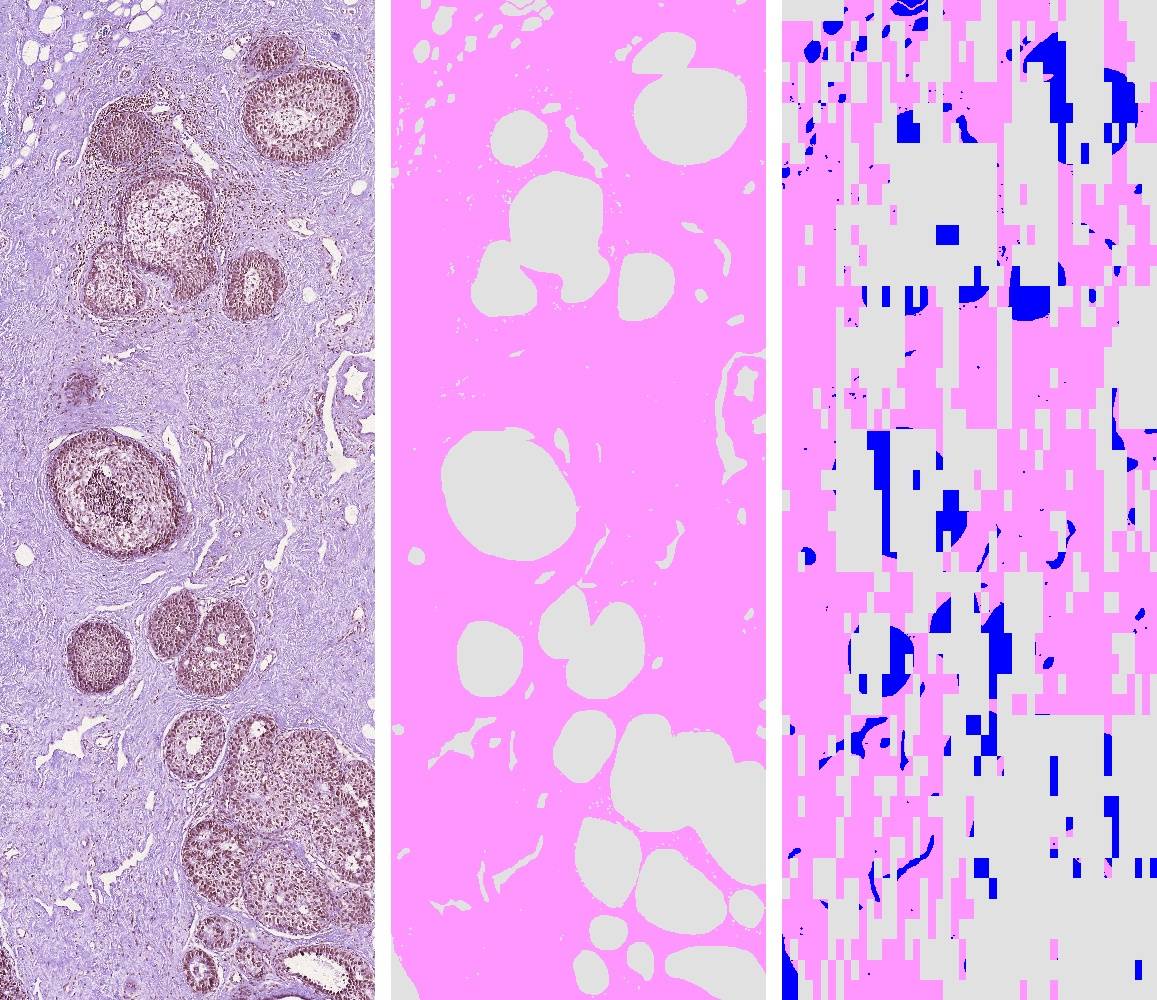}
        \caption{}
        \label{fig:stromal_analysis_a}
    \end{subfigure}
    \hfill
    \begin{subfigure}[b]{0.48\columnwidth}
        \centering
        \includegraphics[height=70px]{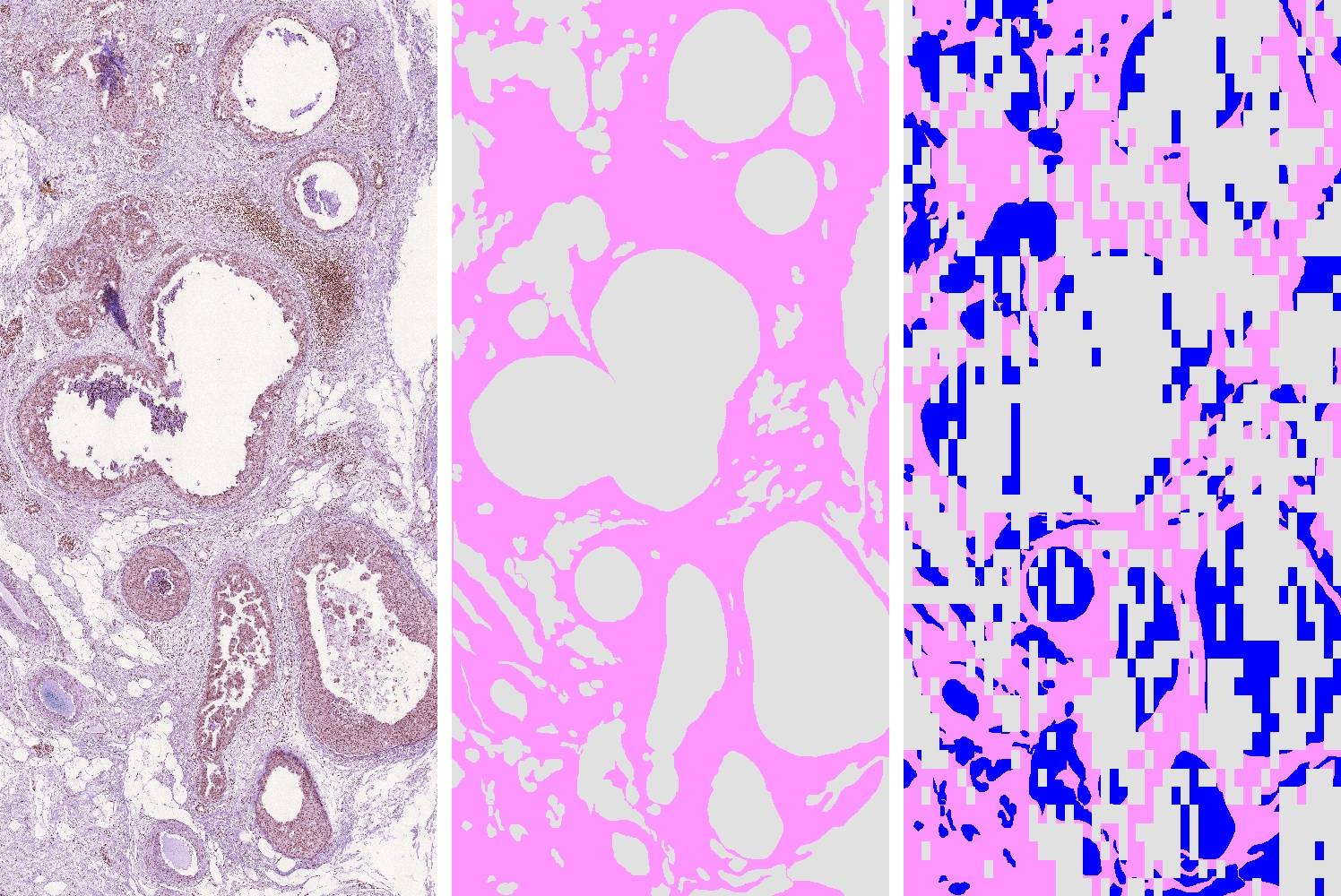}
        \caption{}
        \label{fig:stromal_analysis_b}
    \end{subfigure}
    \hfill
    \begin{subfigure}[b]{0.5\columnwidth}
        \centering
        \includegraphics[height=70px]{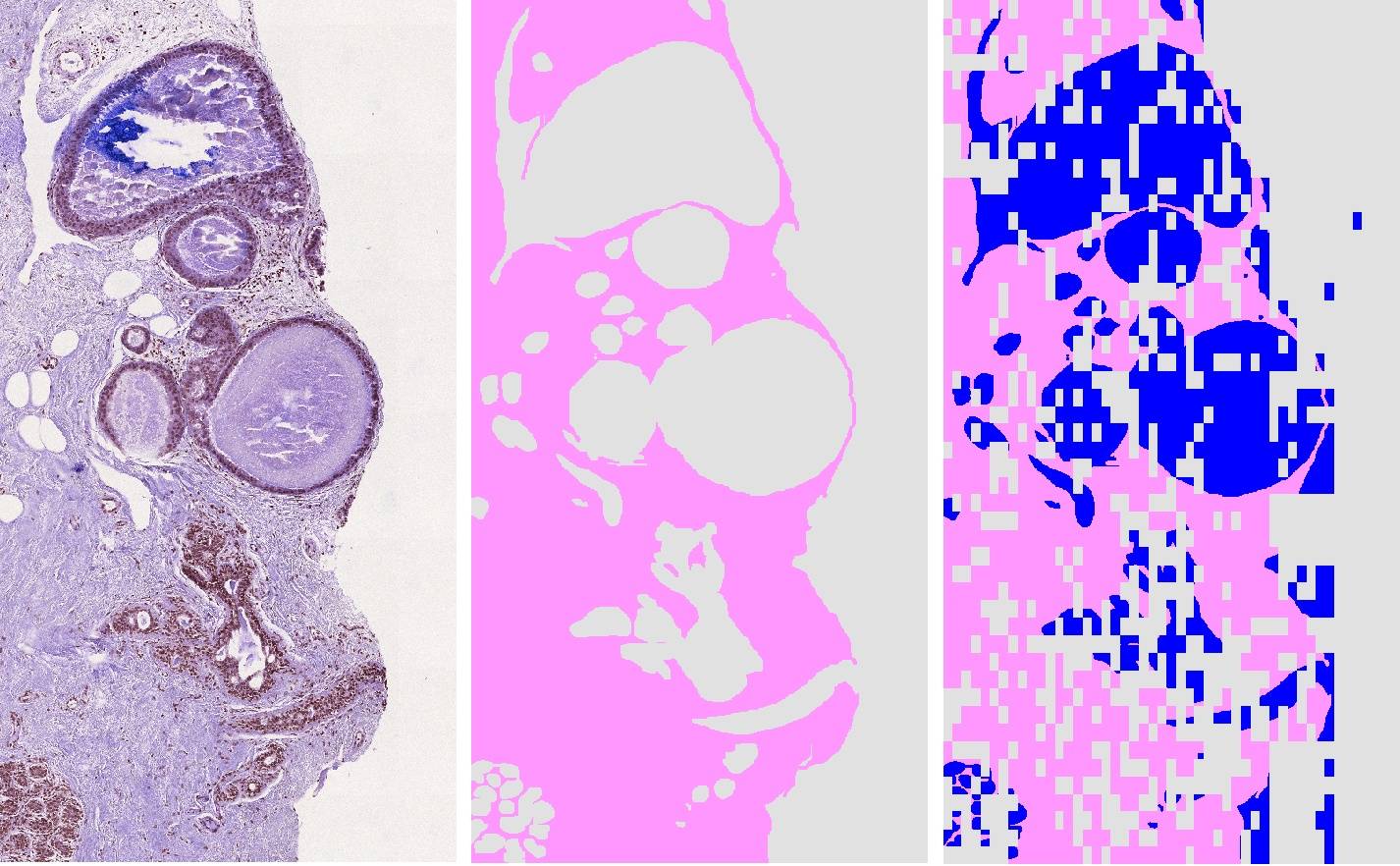}
        \caption{}
        \label{fig:stromal_analysis_c}
    \end{subfigure}
    \hfill
    \begin{subfigure}[b]{0.54\columnwidth}
        \centering
        \includegraphics[height=70px]{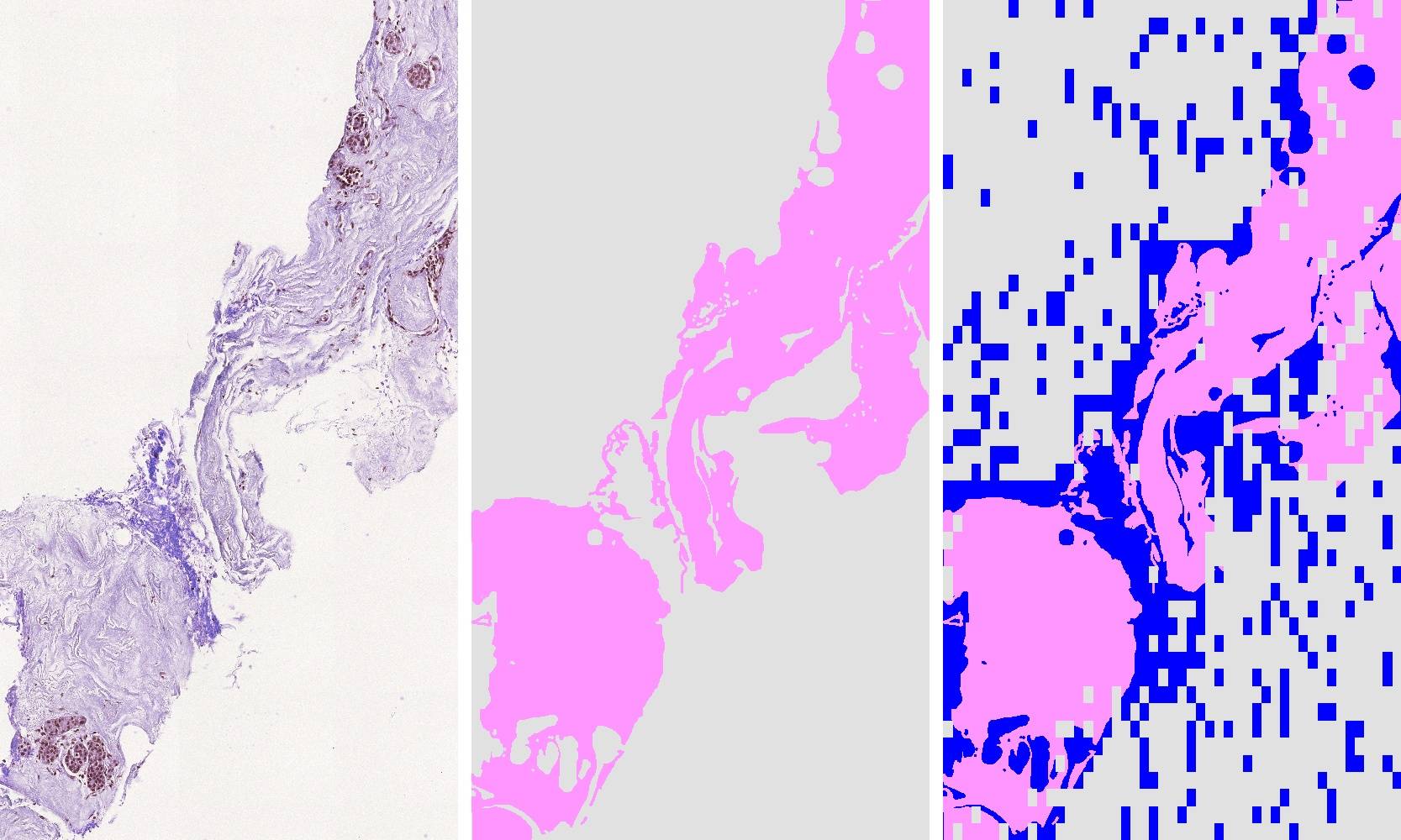}
        \caption{}
        \label{fig:stromal_analysis_d}
    \end{subfigure}
    \vfill
    \begin{subfigure}[b]{0.58\columnwidth}
        \centering
        \includegraphics[width=\columnwidth]{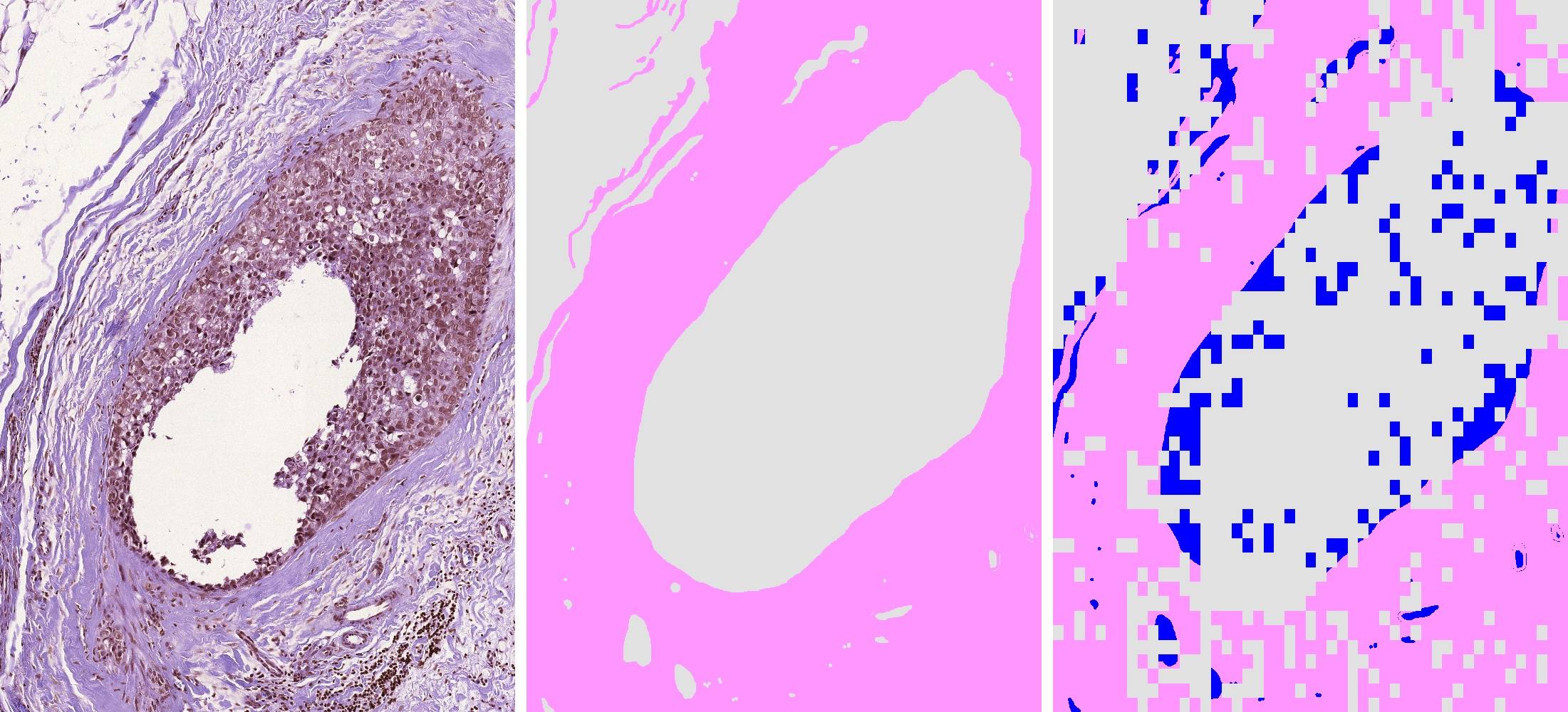}
        \caption{}
        \label{fig:stromal_analysis_e}
    \end{subfigure}
    \hfill
    \begin{subfigure}[b]{0.7\columnwidth}
        \centering
        \includegraphics[width=\columnwidth]{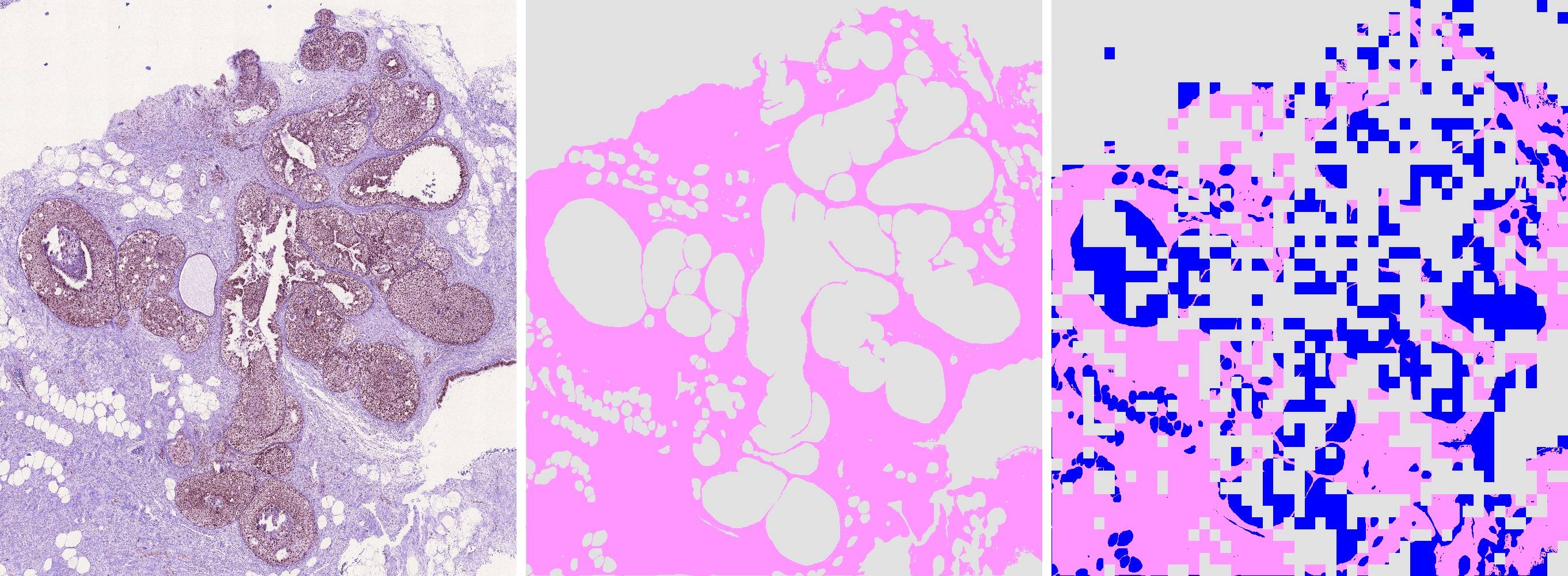}
        \caption{}
        \label{fig:stromal_analysis_f}
    \end{subfigure}
    \hfill
    \begin{subfigure}[b]{0.65\columnwidth}
        \centering
        \includegraphics[width=\columnwidth]{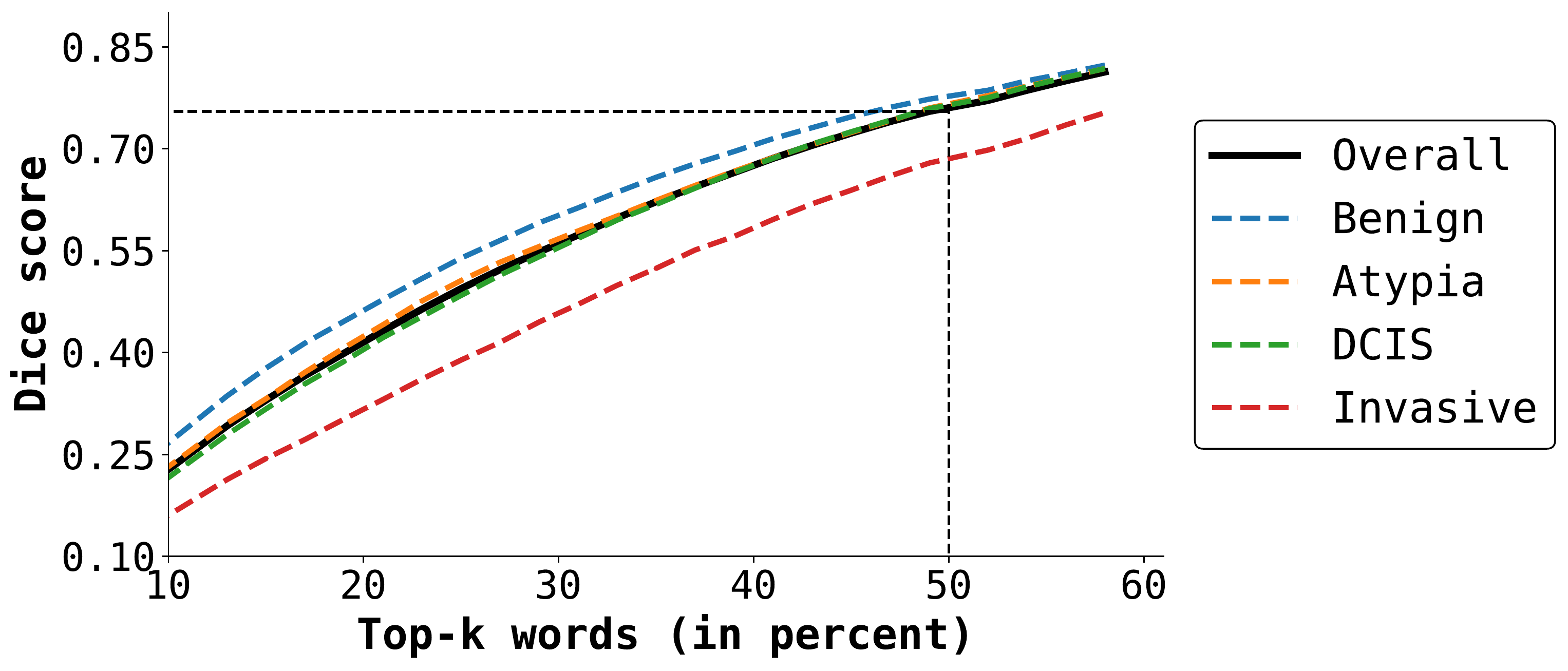}
        \caption{}
        \label{fig:stromal_analysis_g}
    \end{subfigure}
    \caption{\arch~identifies stroma as an important tissue. In (a-f), each sub-figure is organized from left to right as: breast biopsy image, \textcolor{stromaword!150}{\bf  stroma tissue} labeled by pathologists, and the top-50\% words (words that belong to stroma tissue are shown in \textcolor{stromaword!150}{\bf pink} while the remaining words are shown in \textcolor{otherword}{\bf blue}) identified using our model. The remaining 50\% words are shown in white. In (g), we plot the dice score between stromal tissue and top-k word predictions (k varies from 10\% to 60\%) for different diagnostic classes.}
    \label{fig:stromal_analysis}
\end{figure*}

\section{Analysis}
\label{sec:discussion}
Several clinical studies have shown that ductal regions and stromal tissue are important bio-markers for diagnosing breast cancer (e.g., \cite{kinne1989breast,page1998ductal,arendt2010stroma,zhang2012guidelines,conklin2012stroma,conklin2012stroma,mao2013stromal,shah2016management,plava2019recent,desantis2019breast}). Briefly, ducts are thin tubes in the breast and are responsible for carrying milk from lobules (milk glands) to the nipples. These regions are useful in identifying cancers that began in milk ducts (e.g., DCIS \cite{kinne1989breast,page1998ductal,shah2016management}) . On the other hand, the stroma is part of the breast tissue with a structural and developmental role and may be involved in tumor promotion and progression. Many clinical studies have underlined the importance of stroma in tumor progression along with its contribution to risk factors that determines tumor formation \cite{arendt2010stroma,conklin2012stroma}.

It was shown above that our model learns better representations, resulting in significant performance gains compared to existing methods (Table \ref{tab:comparison_sota}). A closer analysis (see Figure \ref{fig:duct_analysis} and \ref{fig:stromal_analysis}) reveals that our model pays attention to these important bio-markers, which helps it to achieve these gains. 

\vspace{1mm}
\noindent \textbf{Ductal regions:} To evaluate if our model pays attention to ductal regions or not, we compute the overlap between ductal regions (marked by experts) and top-k bag predictions of our model using dice score\footnote{We are interested in evaluating if our model pays attention to ductal regions or stroma region. Therefore, we only use top-k bags or words inside these regions while computing the dice score.}. Results are shown in Figure \ref{fig:duct_analysis_g}.  When considering top-50\% bag predictions, \arch~achieves a dice score of 0.68. This shows that \arch~identifies ductal regions as an important structure. Furthermore, Figure \ref{fig:duct_analysis_a}-\ref{fig:duct_analysis_f} shows that \arch~is able to differentiate between ducts of variable size and texture. This shows that \arch~is effective in modeling inter-bag and inter-word relationships, which helps it improve performance.

\vspace{1mm}
\noindent \textbf{Stromal tissue:} We compute the overlap between stromal tissue (pixel-level annotations by pathologists) and top-k word predicted by our model to determine whether our model pays attention to stromal tissue. We use dice score to measure the overlap and vary k from 10\% to 60\%. Figure \ref{fig:stromal_analysis_g} shows that \arch~achieves a dice score of about 0.75 when top-50\% word predictions are considered. This indicates that \arch~also identifies stroma as an important tissue. This is further strengthened by visualizations in Figures \ref{fig:stromal_analysis_a}-\ref{fig:stromal_analysis_f}, which shows that majority of top-50\% words lie in stromal tissue.

\section{Conclusion}
\label{sec:conclusion}
This paper introduces an end-to-end attention-based network for classifying breast biopsy images. \arch~extends bag-of-words models using transformers \cite{vaswani2017attention} to learn global representations. Our approach effectively aggregates inter-word and inter-bag representations, which helps the model to learn representations from clinically relevant tissue structures. \arch~improves the state-of-the-art significantly and matches the classification performance of pathologists on the test set. Furthermore, \arch~learns representations from clinically relevant structures. We believe that top words and bags identified using our method will help us build tools to annotate cell-level structures (e.g., mitoses), which would help in providing detail explanation of diagnosis decisions. In the future, we plan to build such tools and also apply \arch~to other histopathological images, including melanoma.

\bibliographystyle{IEEEtran}  
\bibliography{main}

\end{document}